\documentclass[12pt,a4paper]{article}
\usepackage{lipsum}
\usepackage{amsfonts}
\usepackage{epstopdf}
\usepackage{xcolor}
\usepackage{mathtools}
\usepackage{amssymb}
\usepackage{esint}
\usepackage{bm}
\usepackage{array}
\usepackage{float}
\usepackage{caption}
\usepackage{subcaption}
\usepackage{enumerate}
\usepackage{algpseudocode}
\usepackage{algorithm}
\usepackage{listings}
\usepackage{geometry}
\usepackage{tabularx}
\usepackage[hidelinks]{hyperref} 
\hypersetup{colorlinks=True,
            citecolor=blue} 
\usepackage{placeins}
\usepackage{multirow}
\usepackage{authblk} 
\usepackage{graphicx} 
\usepackage{setspace}
\usepackage{mathtools}
\usepackage{lineno}

\newcommand{\bracfrac}[2]{\left( \frac{#1}{#2} \right)}
\newcommand{\bracked}[1]{\left( #1 \right)}

\usepackage[style = apa, uniquename=false]{biblatex}
\DeclareLanguageMapping{english}{english-apa}
\let\cite\parencite

\let\vec\mathbf

\usepackage{geometry}
\title{On the numerical representation of atmospheric Lamb waves in dynamical cores}
\author{Timothy C. Andrews, Nicholas C. Forcone, Nicholas H. Androski, Owen K. Hughes, Christiane Jablonowski}
\affil{Department of Climate and Space Sciences and Engineering, University of Michigan, Ann Arbor, MI, USA}

\begin{document}

\maketitle

\begin{abstract}
Atmospheric Lamb waves are horizontally propagating acoustic waves that are supported by hydrostatic and nonhydrostatic compressible equation sets. They are classically defined in an isothermal atmosphere and generate no vertical velocity. In a dynamical core, the numerical representation of the Lamb wave is sensitive to modelling choices, such as the upper boundary condition and spatial discretisation. To evaluate the numerical Lamb wave in dynamical cores, we present a new idealised test case that contains a localised pressure perturbation to primarily excite the Lamb wave mode, in a nonrotating, isothermal atmosphere at rest. Lamb waves are identified through their phase speed and the vertical structure of their zonal velocity in three dynamical cores: Spectral Element (SE), Finite-Volume Cubed-Sphere (FV3), and Model for Prediction Across Scales (MPAS), which are available in the Community Atmosphere Model (CAM). Consistent with vertical structure equation theory, the pressure-based coordinate models with an isobaric lid (SE, FV3) generate reduced Lamb wave speeds with low model tops; we show speed reductions of around $25\%$ for a 10 km model top and an isothermal initial temperature of 275 K. For a rigid lid upper boundary condition, such as for the height coordinate MPAS model, the Lamb wave speed is virtually independent of the model top height. Wave amplitudes differ substantially among the tested configurations, with SE producing the largest amplitudes. MPAS' time-adjusted acoustic filter strongly attenuates and broadens the Lamb wave signal, whilst the amplitude and propagation speed are sensitive to the dynamics time step in FV3. Zonal wavenumber-frequency spectra reveal how well the numerical dispersion relations of Lamb and gravity waves follow the theoretical relations, and show element-order-dependent behaviour in SE consistent with linear dispersion theory. The test provides a simple framework for evaluating the effects of the upper boundary conditions, damping, and dispersion in atmospheric dynamical cores.
\end{abstract}

\clearpage

\section{Introduction}
Of all the waves supported by the governing equations for atmospheric flow, one could argue that the Lamb wave is the most overlooked. The Lamb wave is an acoustic mode that propagates exclusively in the horizontal plane, radiating outward from a disturbance, and is supported by compressible hydrostatic and nonhydrostatic equation sets. Atmospheric Lamb waves are external modes, which means they have an exponential variation in the vertical direction, as opposed to the typical sinusoidal pattern. The earliest literature defines the Lamb wave as a normal mode that has no vertical velocity and is supported in an isothermal atmosphere; we will call this the \textit{classical} Lamb wave. However, an analogous mode exists for more realistic (nonisothermal) temperature profiles, and can exhibit small vertical velocities. In this work, we will focus on the representation of the Lamb wave in a dynamical core, which we will term the \textit{numerical} Lamb wave. \par
The first study of classical Lamb waves is attributed to Sir Horace Lamb \parencite{lamb1911atmospheric,lamb1932hydrodynamics}, who investigated normal modes on the sphere and identified the horizontally propagating acoustic mode. Later contributions to atmospheric Lamb wave theory include \textcite{garrett1969atmospheric}, \textcite{bretherton1969lamb}, and \textcite{lindzen1972lamb}, with investigations of the vertical profile of Lamb waves and their characteristics in realistic atmospheres. In the Earth's atmosphere, Lamb waves can be detected after events that generate sufficiently large pressure disturbances, such as volcanic eruptions or meteorite impacts. One of the first observational studies of Lamb wave emission was the Krakatoa eruption of 1883 \parencite{symons1888eruption}. A recent source of strong atmospheric Lamb waves was the 2022 Hunga Tonga--Hunga Ha'apai volcanic eruption; these Lamb waves were examined numerically in studies such as \textcite{matoza2022atmospheric}, \textcite{zhang2025dispersion}, \textcite{amores2022numerical}, \textcite{kulichkov2022acoustic}, including modelling propagation into the ionosphere in \textcite{liu2023atmospheric}. Other studies have examined Lamb wave generation from artificial sources, such as tropospheric nuclear weapons testing in the 1940s--1960s \parencite{donn1962atmospheric,donn1967exploring}. \par
In the dynamical core literature, the Lamb wave mode is typically folded into the categories of external modes or fast waves, rather than being discussed on its own. Fast waves travelling at or near the speed of sound can inhibit the use of explicit time stepping methods, as they lead to severe limits on the allowable time step size. Fast waves are often viewed as physically insignificant in an atmospheric model, e.g. Chapter 8 of \textcite{Durran}, and are typically handled in two ways. The first is the use of implicit techniques, such as the semi-implicit semi-Lagrangian method \parencite{robert1982semi,staniforth1991semi,ritchie1995implementation}, implicit-explicit (IMEX) schemes \parencite{pareschi2000implicit,colavolpe2017rk,weller2013runge}, or linear or fully implicit time stepping \parencite{cotter2023compatible, lee2024comparison}. The second approach is the use of filtered equations, such as those generated by the Boussinesq, anelastic \parencite{ogura1962scale,lipps1982scale,bannon1996anelastic}, or pseudo-incompressible \parencite{durran1989improving,klein2009asymptotics} approximations, or the soundproof model of  \textcite{arakawa2009unification}, which retains compressibility and Lamb waves whilst filtering the vertically propagating acoustic waves \parencite{qaddouri2021implementation,klein2016doubly}. We note that the hydrostatic approximation partially filters the fast waves, as its enforcement of hydrostatic balance prohibits the vertical propagation of sound waves, whilst the Lamb wave is present in both hydrostatic and nonhydrostatic compressible systems. \par
An additional mechanism that can damp Lamb waves in dynamical cores is implicit or explicit dissipation. This may occur through general dissipation mechanisms or filters to keep the dynamical core numerically stable, or the presence of a specific operation that targets the Lamb waves. An example of the latter is ``external mode damping'', which applies a diffusive operator, such as the Laplacian, to the vertically integrated horizontal divergence \parencite{washington1975method,klemp2007conservative,jablonowski2011pros}. The prevalence of dissipative processes means that Lamb waves are rarely observed in idealised or more realistic dynamical core tests. An exception is \textcite{hughes2023mountain}, which noted much stronger Lamb waves in a spectral element dynamical core compared to other models in a baroclinic wave test with topography. \par
We present a simple test case to compare the numerical representation of Lamb waves between dynamical cores. Test cases are widely used in the dynamical core literature to examine the modelling of many types of waves, including gravity waves \parencite{skamarock1994efficiency,baldauf2013analytic}, acoustic waves \parencite{baldauf2014analytical}, baroclinic waves \parencite{jablonowski2006baroclinic, ullrich2014proposed,hughes2023mountain}, and mountain waves \parencite{andrews2026mountain,schar2002new}. However, we know of no recorded test cases designed to evaluate the Lamb wave. Our simple initial condition applies a perturbation to the initial pressure field in an isothermal, nonrotating atmosphere at rest and without orography. \par
Our Lamb wave test case is applied to three dynamical cores, run through the Community Atmosphere Model version 6.4 (CAM) from the National Science Foundation's (NSF) National Center for Atmospheric Research (NSF NCAR): (a) the Spectral Element (SE) model \parencite{lauritzen2018ncar}, (b) the Finite-Volume Cubed-Sphere dynamical core (FV3) from NOAA's Geophysical Fluid Dynamics Laboratory (GFDL) \parencite{harris2021scientific_FV3_doc,lin1996multidimensional,lin2004vertically}, and (c) the Model for Prediction Across Scales (MPAS) \parencite{skamarock2012multiscale,skamarock2025description}. We first confirm that Lamb waves are generated in all models via two key characteristics: (i) the expected phase speeds for a range of isothermal temperatures and (ii) the expected exponential increase in horizontal wind velocity with height. We show that the upper boundary condition has a significant impact on the numerical Lamb waves. In line with the theory of \textcite{de2020barotropic}, models with a pressure coordinate and an isobaric lid, which enforces a vanishing vertical pressure velocity at the upper boundary, lead to reduced wave speeds with lower model tops (SE, FV3). The height coordinate model (MPAS) with a rigid lid that enforces a vanishing vertical velocity $w$ at the upper boundary has no speed change with model top height. \par
Variations in Lamb wave amplitude between the models are observed as a consequence of modelling features, including MPAS' time-adjusted acoustic filter, which specifically targets the Lamb wave, and FV3's timestepping scheme, which damps divergent motions and has an accuracy that is sensitive to the dynamics time step size. The Lamb wave test also allows the analysis of linear dispersion relation accuracy; we show this through the order of the SE basis functions impacting its numerical dispersion relation. Overall, the Lamb wave test provides a simple framework to examine the effect of boundary conditions, dispersion, and linear wave propagation in dynamical cores.\par
The layout of the rest of the paper is: Section \ref{sec:lamb_wave_theory} describes the theory for classical and numerical Lamb waves, from the governing equations and vertical structure equation. Section \ref{sec:dynamical_cores} overviews the three dynamical cores we examine, and the Lamb wave test case is outlined in Section \ref{sec:test_case}. Simulations with the dynamical cores are then presented, with a validation of the Lamb waves (Section \ref{sec:validation}) and a comparison of models and settings (Section \ref{sec:sensitivites}). Finally, Section \ref{sec:conclusions} summarises the main results and highlights our future research directions. \par

\section{Atmospheric Lamb wave theory}
\label{sec:lamb_wave_theory}
We begin with relevant theoretical results for classical and numerical Lamb waves. The classical Lamb wave is identified through the governing equations, whilst the numerical Lamb wave is described by an eigenmode analysis of the vertical structure equation. We then compare the dispersion relation and speed of the Lamb wave to those of the discretised gravity waves. Note that we use the term \textit{speed} to refer to the phase speed throughout. 

\subsection{Derivation from governing equations}
\label{subsec:govern_equations}
The classical Lamb wave is identified in the governing compressible equations for fluid flow in the Earth's atmosphere after two key assumptions: i) an isothermal atmosphere, and ii) the absence of vertical velocity, e.g. \textcite{satoh2013atmospheric}, \textcite{Vallis}, and \textcite{gill1982atmosphere}. Neglecting the Earth's rotation and assuming a hydrostatic, inviscid atmosphere, the Lamb wave dispersion relation is shown to be
\begin{equation}
    \omega = \pm c_{\text{L}} k_{\text{h}},
\label{eq:lamb_disp_rel}
\end{equation}

\noindent with $k_{\text{h}}$ the horizontal wavenumber on the sphere. This dispersion relation describes the angular frequency $\omega$ of nondispersive acoustic waves that travel horizontally at speed $c_{\text{L}}$ and have no dependence on the vertical wavenumber $m$. The classical Lamb wave speed, which is equal to the speed of sound in an isothermal atmosphere of temperature $T_0$, is the constant
\begin{equation}
    c_{\text{L}} = \sqrt{\gamma R_{\text{d}} T_0},
\label{eq:lamb_speed}
\end{equation}

\noindent where $\gamma=c_{\text{p}}/c_{\text{v}}=7/5$ for a diatomic ideal gas, with $c_{\text{p}}=1004.64 ~\text{J} ~\text{kg}^{-1} ~\text{K}^{-1} $ and $ c_{\text{v}}$ the specific heat capacities at constant pressure and volume, respectively, and $R_{\text{d}} = 287.04 ~\text{J} ~\text{kg}^{-1} ~\text{K}^{-1}$ is the specific gas constant for dry air. \par 
We note that when considering planetary rotation, the Lamb wave is no longer nondispersive, as it includes a contribution from the Coriolis force. However, the nonrotating Lamb wave frequency is generally much larger than a characteristic value of the Coriolis parameter, so the dispersion relation is typically well approximated by (\ref{eq:lamb_disp_rel}). To simplify the identification and analysis of Lamb waves in a dynamical core, our test case will neglect the Earth's rotation. \par
Whilst the classical Lamb wave behaves as a compressional wave in the horizontal dimensions, it has an exponential structure in the vertical dimension. The horizontal velocity perturbation associated with the classical Lamb wave increases with height, such as for the zonal wind perturbation ($u_{\text{L}}^{'}$),
\begin{equation}
    u_{\text{L}}^{'} \propto \exp\bracfrac{\kappa z}{H} = \exp\bracfrac{2z}{7H},
\label{eq:lamb_u_vert_dependence}
\end{equation}

\noindent with $\kappa=R_{\text{d}}/c_{\text{p}} = 2/7$, and $H$ is the isothermal scale height, 
\begin{equation}
    H = \frac{R_{\text{d}} T_0}{g},
\label{eq:iso_scale_height}
\end{equation}

\noindent with $g = 9.80616 ~\text{m} ~\text{s}^{-2}$ representing Earth's gravity. The pressure perturbation of the classical Lamb wave decreases with altitude at a rate of $\exp(-5z/(7H))$. The Lamb wave kinetic energy density decays as $\exp(-3z/(7H))$ in the vertical \parencite{Vallis,bretherton1969lamb}. \par
In this work, we will use the terms `internal' and `external' from a wave propagation perspective, e.g. \textcite{satoh2013atmospheric}, where the vertical structure is proportional to $\exp(imz)$. This defines an internal wave as having a real vertical wavenumber ($m^2 > 0$) whilst an external wave has an imaginary vertical wavenumber ($m^2 <0$). The classical Lamb wave is an external mode due to the exponential dependence on height (\ref{eq:lamb_u_vert_dependence}). \par

\subsection{Vertical eigenstructure}
\label{subsec:eigenstuff}
An alternative approach for identifying atmospheric Lamb waves is through the structure equations \parencite{gill1982atmosphere,lindzen1969atmospheric}. This incorporates the effect of boundary conditions, which is critical for specifying the numerical Lamb waves that are supported by a dynamical core. \par
The structure equation approach invokes a separation of variables to form horizontal and vertical equations that are coupled through the equivalent depth, $h_n$, where the mode number $n$ is a non-negative integer. The vertical structure function, $L_n(z)$, is described by a Sturm-Liouville problem that can be solved analytically for a constant temperature profile, or using numerical methods for nonisothermal atmospheres \parencite{de2020barotropic,de2024effect,castanheira1999numerical}. The general solution in an isothermal atmosphere (and $h_n \neq 4\kappa H$) is
\begin{equation}
    L_n(z) = e^{\frac{z}{2H}} [C_1 e^{i m_n z} + C_2 e^{-i m_n z}],
\label{eq:gen_sturm_sol}
\end{equation}

\noindent where $C_1$ and $C_2$ are constants, and the vertical wavenumber for mode $n$ is given by
\begin{equation}
    m_n = \frac{1}{2H} \sqrt{\frac{4\kappa H}{h_n} - 1}.
\label{eq:vertical_waveno}
\end{equation}

\noindent The speed of each eigenmode, $c_n$, is determined from the equivalent depth $h_n$, by
\begin{equation}
    c_n=\sqrt{gh_n}.
\end{equation}

The forms of $C_1$ and $C_2$, and thus the specific solution to the vertical structure equation, are determined from the boundary conditions. We consider the case of no vertical velocity, $w=0$, at the bottom boundary. For the upper boundary, we consider two cases:
\begin{enumerate}
    \item The \textit{rigid lid}: $w=0$ at the model top. This condition enforces $Dz/Dt = w = 0$, where $D/Dt$ is the material derivative, which is natural for a height-based vertical coordinate. The $w=0$ condition can be expressed as a Robin boundary condition on $L_n(z)$.
    \item The \textit{isobaric lid}: $L_n(z)=0$ at the model top. As the vertical structure function is proportional to the vertical pressure velocity $\omega_p$\footnote{Note the subscript $p$ to distinguish the vertical pressure velocity $\omega_p$ from the angular frequency $\omega$.}\parencite{lindzen1969atmospheric}, this choice enforces $Dp/Dt = \omega_p = 0$, and so is a natural choice for a pressure-based vertical coordinate. The isobaric lid permits a nonzero $w$ at the model top.
\end{enumerate}

\noindent Note that \textcite{de2020barotropic} also considers a boundary condition of $L_n'(z)=0$, which at the upper boundary approximates a radiative boundary for sufficiently high model tops. We do not consider this, as it is not a common option in dynamical cores. \par
We now describe how the choice of a rigid or isobaric lid impacts the numerical Lamb wave. We use $z_{\text{T}}$ to denote the model top height.

\subsubsection{The rigid lid}
The application of $w=0$ at both the bottom and top boundaries permits an analytical solution for $L_n(z)$ \parencite{de2020barotropic,castanheira1999numerical}. The $n=0$ mode, which is classified as the barotropic mode, has an equivalent depth that is independent of the model top height,
\begin{equation}
    h_0 = \gamma H.
\label{eq:eq_depth}
\end{equation}

\noindent This means that $c_0 =\sqrt{gh_0} = \sqrt{\gamma R_{\text{d}}T_0}$, so the speed of the $n=0$ mode matches that of the classical Lamb wave (\ref{eq:lamb_speed}). The vertical wavenumber (\ref{eq:vertical_waveno}) satisfies $m^2<0$, so this mode is external. In the absence of orography, the vertical structure function of the external barotropic mode is 
\begin{equation}
   L_0(z) = A_0 \exp \bracked{\frac{\kappa z}{H}},
\label{eq:L0_vert_profile}
\end{equation}

\noindent where $A_0$ is an amplitude. Thus, $L_0(z)$ matches the 2/7 exponential vertical structure of the classical Lamb wave (\ref{eq:lamb_u_vert_dependence}). \par

\subsubsection{The isobaric lid}
For an upper boundary condition of $L_n(z)=0$, the solutions to the vertical structure equation are semi-analytic and can be obtained using root-finding methods; see Appendix \ref{append:cL_p_coord} for further details. \textcite{de2020barotropic} showed that for these boundary conditions, the fastest mode has a speed that is a function of the layer thickness. For a numerical model, this implies a dependence on the model top height, $z_{\text{T}}$. We will use asterisks to denote quantities associated with the isobaric lid, specifically an \textit{effective} equivalent depth $h^*$ and speed $c^* = \sqrt{gh^*}$ that depend on $z_{\text{T}}$.  \par
A critical impact of an effective equivalent depth that varies with model top height is that the vertical wavenumber $m^*$ of the numerical Lamb wave also varies with $z_{\text{T}}$. For $z_{\text{T}} > 8H/3, h^* > 4 \kappa H$ and $(m^*)^2 < 0$, so the Lamb wave is external. However, the wave is internal when $z_{\text{T}} < 8H/3$, $h^* < 4 \kappa H$, and $(m^*)^2 > 0$. In the nomenclature of \textcite{de2020barotropic}, the numerical Lamb wave is the barotropic mode for $z_{\text{T}} > 8H/3$ and the $n=0$ baroclinic mode for $z_{\text{T}} < 8H/3$. The numerical Lamb wave speed smoothly transitions between these regimes, with $c^*=2NH$ at the critical height of $z_{\text{T}} = 8H/3$; here, $N$ is the Brunt-V{\"a}is{\"a}l{\"a} frequency that governs static stability of the atmosphere, and for an isothermal atmosphere is the constant 
\begin{equation}
    N = \sqrt{\frac{g^2}{c_{\text{p}} T_0}}.
\label{eq:brunt_vais_freq}
\end{equation} \par
\begin{figure}
    \centering
    \includegraphics[width=0.5\linewidth]{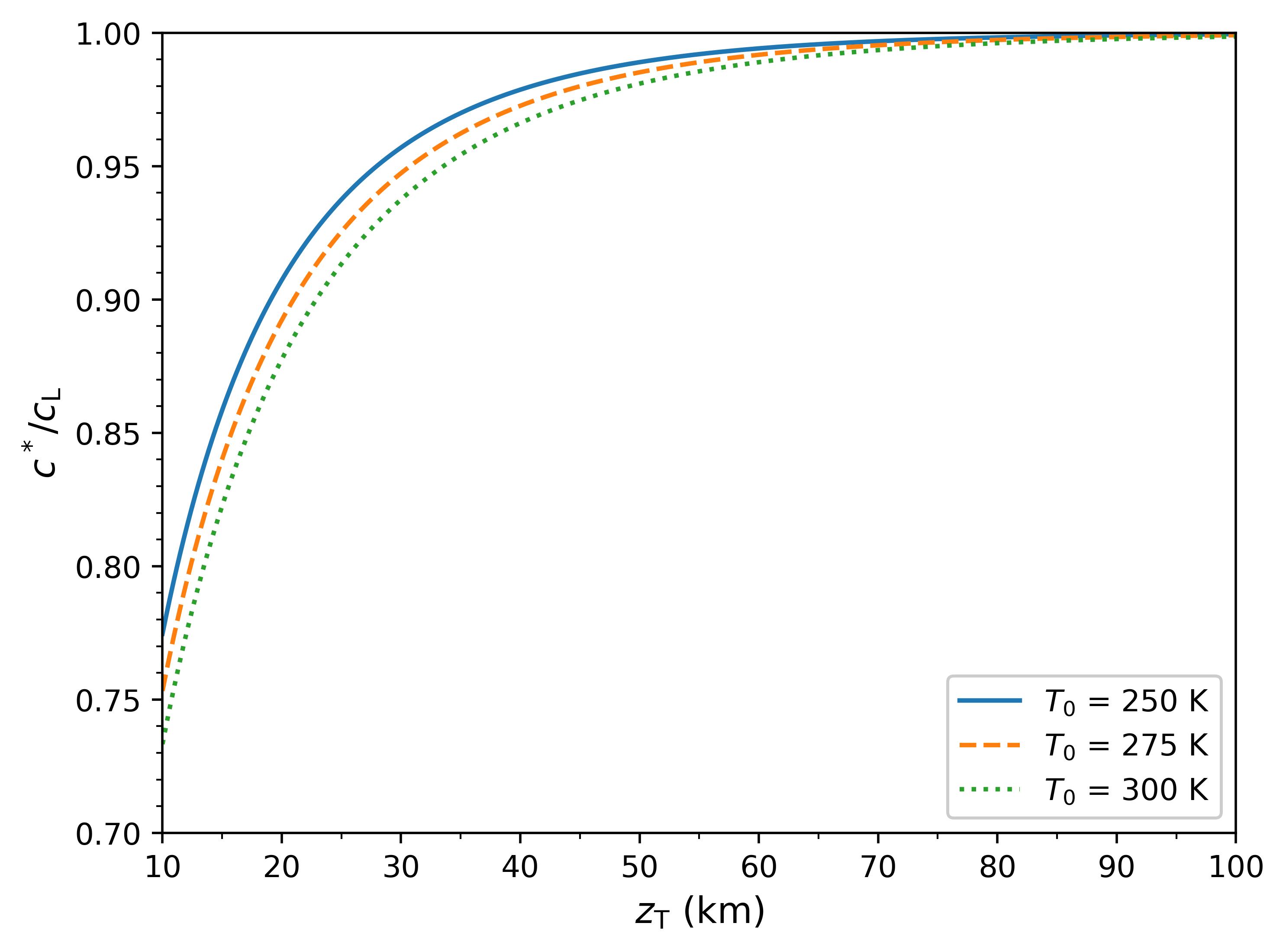}
    \caption{The ratio of the isobaric lid Lamb wave speed ($c^*$) to the classical Lamb wave speed ($c_{\text{L}}$) with model top height $z_{\text{T}}$, for different isothermal temperatures $T_0$. }
    \label{fig:lamb_wave_ratios}
\end{figure}

In the limit of $z_{\text{T}} \rightarrow \infty$, the speed of the isobaric lid numerical Lamb wave approaches that of the classical Lamb wave. However, for a finite model top, as is the case in any numerical simulation, we have $c^* < c_{\text{L}}$ and a slower Lamb wave with an isobaric lid compared to the rigid lid. Figure \ref{fig:lamb_wave_ratios} shows the reduction in wave speed with the isobaric lid by plotting the ratio of $c^*/c_{\text{L}}$. With a high model top of 100 km, $c^*/c_{\text{L}} = 0.999$ and the numerical Lamb wave speed with the isobaric lid is virtually identical to that of the classical Lamb wave. However, a low model top substantially slows the isobaric lid Lamb wave: for an isothermal temperature of $T_0=275$ K, the numerical Lamb wave speed is 89\% of the classical speed for a 20 km model top and 75\% of the classical counterpart for a 10 km model top. The speed reductions with the isobaric lid are exacerbated by an increased temperature (Figure \ref{fig:lamb_wave_ratios}). \par

To summarise the analysis of the vertical structure equation in this section, we expect that the numerical Lamb wave speed will be independent of $z_{\text{T}}$ for a rigid lid, but significantly slowed for the isobaric lid with low model tops. This expectation will be examined in the numerical tests with the SE, FV3, and MPAS dynamical cores (Sections \ref{sec:validation} and \ref{sec:sensitivites}). \par

\subsection{Differentiating external Lamb and internal gravity waves}
We now discuss the gravity waves supported by a numerical model and how these can be distinguished from the numerical Lamb wave.
Practically, the gravity wave speeds in a numerical model depend on a number of choices such as the model top height, upper boundary condition, and choice of vertical grid staggering \parencite{de2020barotropic,derrida2026propagation}. The fastest gravity wave for each mode $n$ occurs in the long-wave limit of $k_{\text{h}} \rightarrow 0$,
\begin{equation}
    c_{n, \text{lw}}= \frac{2NH}{\sqrt{1 + 4m_n^2H^2}},
\label{eq:grav_speeds}
\end{equation}

\noindent with $n \geq 1$ (see Appendix \ref{append:gravity_wave_speed} for a derivation of (\ref{eq:grav_speeds})). For a rigid lid, the vertical wavenumbers are
\begin{equation}
    m_n = \frac{n\pi}{z_{\text{T}}}.
\label{eq:normed_vert_wavenumbers}
\end{equation}

\noindent For an isobaric lid, the analogous long-wave limit gravity wave speeds $c_{n, \text{lw}}^*$ follow (\ref{eq:grav_speeds}) but with an effective wavenumber of $m_n^*$ in place of $m_n$. The effective wavenumbers $m_n^*$ are computed numerically, as described in Appendix \ref{append:gravity_wave_speed}, and vary as a function of the model top height. The long-wave limit gravity wave speeds are slower with the isobaric lid compared to the rigid lid, with $m_n^* > m_n$ so that $ c_{n, \text{lw}}^* < c_{n, \text{lw}}$ for all modes $n \geq 1$. \par
The maximum theoretical gravity wave speed in the atmosphere occurs for $m_n \rightarrow 0$ in the long-wave limit (\ref{eq:grav_speeds}):
\begin{equation}
    c_{\text{grav,max}} = 2 N H =  \frac{2R_{\text{d}}}{\sqrt{c_{\text{p}}}} \sqrt{T_0}.
\label{eq:grav_speed_fastest}
\end{equation}

\begin{figure}
    \centering
    \includegraphics[width=0.5\linewidth]{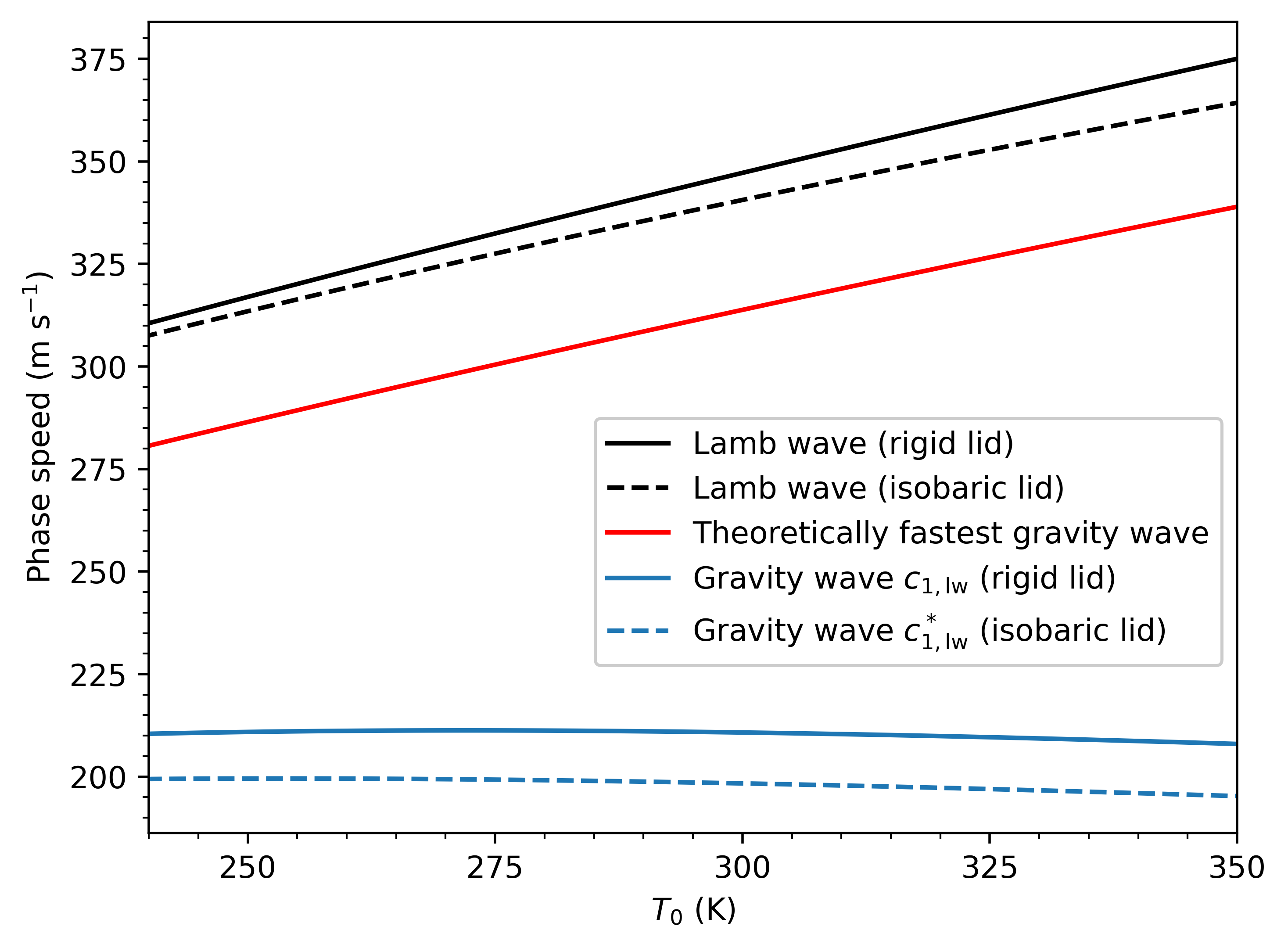}
    \caption{Theoretical Lamb waves and long-wave limit $n=1$ gravity wave speeds for varying isothermal temperature $T_0$, for the rigid and isobaric lids with a model top height at 50 km. The theoretically fastest gravity wave speed is $2NH$.}
    \label{fig:analyt_lamb_grav_speed_with_T0}
\end{figure}

\noindent Thus, the theoretically fastest gravity wave speed is proportional to $\sqrt{T_0}$, as is the case with the classical Lamb wave speed (\ref{eq:lamb_speed}). However, Figure \ref{fig:analyt_lamb_grav_speed_with_T0} shows that there is a significant separation between the classical Lamb and fastest theoretical gravity wave speeds, with $c_{\text{grav,max}}/c_{\text{L}} = 2\sqrt{\kappa/\gamma}\approx 0.904$; this corresponds to about a 32 m s$^{-1}$ difference at 275 K. In the case of a numerical model, the upper boundary condition means that the fastest discretised gravity wave is the $n=1$ mode, which is much slower than the $m_n \rightarrow 0$ gravity wave; the fastest rigid lid gravity wave is $\approx64\%$ of the classical Lamb wave speed for $z_{\text{T}}=50$ km and $T_0=275$ K. \par

Another distinction between the numerical Lamb wave and gravity waves is in the vertical structure. The horizontal velocity perturbation of the gravity waves have a sinusoidal profile in the vertical that is modulated by an exponential growth, for example in the zonal wind perturbation $u_{\text{grav}}^{'}$,

\begin{equation}
    u_{\text{grav}}^{'} \propto \exp\bracfrac{z}{2H},
\label{eq:grav_u_vert_dependence}
\end{equation}

\noindent as highlighted in the general solution to the vertical structure equation (\ref{eq:gen_sturm_sol}). Compared to the $u_{\text{L}}^{'} \propto \exp (2z/(7H))$ structure of classical Lamb waves, the gravity waves have a stronger exponential growth in amplitudes. \par
In the following numerical experiments, the presence of Lamb waves is confirmed by i) the expected phase speeds for a range of isothermal temperatures (Figure \ref{fig:analyt_lamb_grav_speed_with_T0}), and ii) a 2/7 exponential growth in zonal velocity with altitude (\ref{eq:lamb_u_vert_dependence}). Before we do so, we briefly introduce the dynamical cores that are examined. \par

\section{Dynamical cores}
\label{sec:dynamical_cores}
Our Lamb wave test case is applied to three dynamical cores --- SE, FV3, and MPAS --- which are available via the CAM6.4 framework from NSF NCAR. These dynamical cores all use the Lorenz vertical grid staggering approach \parencite{lorenz1960energy,thuburn2005vertical}. Unless otherwise noted, we run the models with an approximate horizontal grid spacing of $1^\circ$ for the results in Sections \ref{sec:validation} and \ref{sec:sensitivites}. The basic characteristics of the dynamical cores are summarised in Table \ref{table:dycores}.

\subsection{SE}
The SE dynamical core is built upon a mimetic spectral element technique \parencite{taylor1997spectral,taylor2010compatible}. It uses an Arakawa A-grid staggering approach \parencite{arakawa1977computational}, with prognostic variables stored at the Gauss-Lobatto-Legendre nodes used for quadrature. By default, cubic basis functions are utilised in each horizontal dimension, which is called the ``np4'' configuration with four collocation points per dimension in each element. An equiangular gnomonic cubed-sphere \parencite{ronchi1996cubed} is used for the horizontal mesh, with a hybrid pressure-based coordinate and floating Lagrangian layers \parencite{lin2004vertically} in the vertical dimension. The $1^\circ$ configuration corresponds to the setting ``ne30np4'' with $30 \times 30$ spectral elements per cubed-sphere panel. We use the hydrostatic version of SE. However, a nonhydrostatic version of SE is also available in development versions of CAM6.4 \parencite{jablonowski2026nsf} which was developed for the Energy Exascale Earth System Model (E3SM) at the U.S. Department of Energy \parencite{taylor2020energy}. The integration of SE into CAM is described in \textcite{lauritzen2018ncar}.

\subsection{FV3}
The FV3 dynamical core was developed by NOAA GFDL \parencite{harris2021scientific_FV3_doc}. In the horizontal, it utilises a finite-volume approach on a cubed-sphere grid that is built upon the \textcite{lin1996multidimensional} transport scheme. In the vertical, the floating Lagrangian algorithm of \textcite{lin2004vertically} is used, which, like SE, uses a hybrid pressure-based vertical coordinate. FV3 uses a unique CD-grid, with the D-grid staggering used to store prognostic variables with interpolations to the C-grid for transport. The CD-grid behaves like the D-grid at leading order \parencite{skamarock2008linear,konor2018impacts}. The FV3 gnomonic cubed-sphere grid uses the equi-edge mapping \parencite{chen2021lmars}, which is similar to the equiangular mapping but with more regularity at panel edges \parencite{andrews2026stability}. We run FV3 in its hydrostatic mode, although there is the option to use a nonhydrostatic solver. The $1^\circ$ configuration is named ``C96'', reflecting its $96 \times 96$ finite volumes per cubed-sphere panel. FV3 typically needs horizontal divergence damping for stability, and we use the sixth-order version in this work. 

\subsection{MPAS}
The finite-volume Model for Prediction Across Scales \parencite{skamarock2012multiscale,skamarock2025description} uses a centroidal Voronoi horizontal mesh \parencite{ju2011voronoi} and a C-grid staggering approach with the TRiSK modification \parencite{ringler2010unified,thuburn2009numerical} for an accurate representation of the Coriolis force. A hybrid height-based vertical coordinate is used, with the option for additional smoothing operators in the presence of steep orography \parencite{klemp2011terrain}. MPAS is a nonhydrostatic model, so the vertical velocity $w$ is a prognostic variable. The $1^\circ$ configuration corresponds to the ``120'' km MPAS grid.\par
A relevant MPAS diffusion mechanism for the modelling of atmospheric Lamb waves is the time-adjusted acoustic filter described in \textcite{klemp2018damping}, also called three-dimensional divergence damping. This mechanism acts on the modified divergence $D_{\theta} = \bm{\nabla} \cdot (\rho_{\text{d}} \theta_{\text{m}} \vec{u})$, with $\rho_{\text{d}}$ the density of dry air, $\theta_{\text{m}}$ a moist potential temperature that is equal to the dry potential temperature in the absence of moisture, and $\vec{u}$ the three-dimensional wind field. The time-adjusted acoustic filter is only applied to the horizontal momentum equation; diffusion in the vertical momentum equation is instead configured via off-centring in the temporal discretisation. The strength of the time-adjusted acoustic filter is governed by the dimensionless parameter $\gamma$ (\texttt{config\_smdiv} in the code) which has a default value of 0.1. We will compare simulations with and without this filter.

\renewcommand{\arraystretch}{1.3}

\begin{table}[htpb]
\caption{Summary of relevant features of the three dynamical cores.}
\centering
\label{table:dycores}
\begin{tabular}{ |p{0.24\linewidth}||p{0.23\linewidth}|p{0.21\linewidth}|p{0.2\linewidth}| } 
\hline
Model aspect & SE & FV3 & MPAS \\ 
\hline\hline
Equation set & Hydrostatic & Hydrostatic & Nonhydrostatic \\
\hline
Horizontal staggering & A-grid & CD-grid & C-grid \\
\hline
Mesh & Equiangular gnomonic cubed-sphere & Equi-edge gnomonic cubed-sphere & Centroidal Voronoi \\
\hline
Vertical staggering & Lorenz & Lorenz & Lorenz \\
\hline
Vertical coordinate & Hybrid pressure & Hybrid pressure & Hybrid $z$ \\
\hline 
Upper boundary \newline condition & Isobaric lid & Isobaric lid & Rigid lid \\
\hline
\end{tabular}
\end{table}

\section{The test case}
\label{sec:test_case}
We now describe the Lamb wave test case and some important aspects of the numerical simulations. We use a spherical coordinate system with the longitude range $\lambda \in [0, 2\pi]$ and latitude range $\phi \in [-\pi/2, \pi/2]$.

\subsection{Initial condition}
The initial condition is reminiscent of test case 1 from \textcite{tomita2004new}, which tests three-dimensional acoustic wave propagation, except that we initialise a localised three-dimensional hydrostatic pressure disturbance to primarily excite Lamb waves. In pressure-based hydrostatic models, this is accomplished by applying a Gaussian perturbation to the surface pressure, 
\begin{equation}
    p_{\text{s}}(\lambda, \phi, t=0) = p_0 + p_{\text{pert}} \exp \bracked{\frac{-r^2}{d^2}},
\label{eq:surf_p}
\end{equation}

\noindent with $p_0=1000 ~\text{hPa}$ a reference pressure, $d$ = 1/30 the radial Gaussian half-width, and $r$ the great circle distance from point ($\lambda_{\text{c}}, \phi_{\text{c}}$) on the sphere of unit radius,
\begin{equation}
    r(\lambda, \phi) = \arccos(\sin(\phi_{\text{c}}) \sin(\phi) + \cos(\phi_{\text{c}}) \cos(\phi) \cos(\lambda_{\text{c}} - \lambda)).
\label{eq:great_arc}
\end{equation}

\noindent The perturbation is centred at $(\lambda_{\text{c}}, \phi_{\text{c}}) = (3\pi/2, 0)$, with the excited Lamb wave travelling to the antipode at $(\lambda_{\text{a}}, \phi_{\text{a}}) = (\pi/2, 0)$ and back. The initial velocities and surface geopotential are zero. A surface pressure perturbation of $p_{\text{pert}} = 50 ~\text{hPa}$ is chosen to have a similar magnitude to the Hunga Tonga--Hunga Ha'apai eruption study of \textcite{amores2022numerical}. Although this defines a large perturbation to the system, the excited dynamics in this test is linear, as further explained in Section \ref{subsec:other_tests}. The initial temperature is set to the spatially constant $T_0$. \par
For models that require the specification of the pressure or density at arbitrary heights, such as at midpoint levels, the pressure at a given height is obtained from the initial surface pressure (\ref{eq:surf_p}) by
\begin{equation}
    p(\lambda,\phi,z,t=0)=p_{\text{s}} \exp \bracfrac{-z}{H}.
\end{equation}

\noindent An initial density can also be specified through an appropriate equation of state. \par
The default parameters for the Lamb wave test case are given in Table \ref{tab:test_case_pars}. These values are used in the numerical experiments, unless stated otherwise. The classical Lamb wave speed with the default parameters, including an isothermal temperature of $T_0 = 275$ K, is 332.4 m s$^{-1}$. \par

\begin{table}[]
\caption{Default parameters for the Lamb wave test case.}
    \centering
    \begin{tabular}{|p{0.13\linewidth}|p{0.34\linewidth}|p{0.2\linewidth}|}
    \hline
        Parameter & Definition & Default value \\
        \hline
        \hline
        $p_0$ & Reference surface pressure & 1000 hPa \\
        \hline
        $p_{\text{pert}}$ & Initial surface pressure perturbation & 50 hPa \\
        \hline
        $T_0$ & Isothermal temperature & 275 K \\
        \hline
        $z_{\text{T}}$ & Reference model top height & 50 km \\
        \hline
        $\Delta z$ & Vertical grid spacing & 500 m \\
        \hline
        $H$ & Isothermal scale height, Eq.  (\ref{eq:iso_scale_height}) & $\approx 8050$ m \\
        \hline 
        $N$ & Brunt-V{\"a}is{\"a}l{\"a} frequency, Eq. (\ref{eq:brunt_vais_freq}) & $\approx 0.0187 ~\text{s}^{-1}$ \\
        \hline
        $\Omega$ & Rotation rate of the Earth & 0 s$^{-1}$ \\
        \hline
    \end{tabular}
    \label{tab:test_case_pars}
\end{table}

\subsection{Simulation considerations}
\label{subsec:sim_considers}
Typically, dynamical cores are run with a sponge layer, which is a region of additional dissipation near the model top that lessens wave reflection from the upper boundary. By default, the three dynamical cores all use a sponge layer with some form of Laplacian diffusion or Rayleigh damping. We run each model with its sponge layer disabled to avoid modifying the vertical structure of the horizontal velocities. Unless otherwise noted, the models are run with their default diffusion parameters. Adiabatic configurations are used to examine the dynamics without any physics parametrisations.  \par
The default $1^{\circ}$ horizontal resolution gives grid spacings of around 100--120 km in the models. The desired output fields (surface pressure, zonal velocity, and geopotential) are interpolated to a regular $1^{\circ} \times 1^{\circ}$ longitude-latitude (lon-lat) grid for post-processing. For SE, the interpolation is performed using its basis functions, whilst the native FV3 and MPAS fields are regridded using bilinear interpolation. \par
The Lamb wave speed is calculated from linear regression on the times at which the maximum surface pressure perturbations reach points along the equator between the centre of the initial disturbance ($\lambda_{\text{c}} = 270^{\circ}$) and the antipode ($\lambda_{\text{a}} = 90^{\circ}$): $\lambda \in \{ 240^{\circ}, 220^{\circ}, 200^{\circ}, 180^{\circ}, 160^{\circ}, 140^{\circ}, 120^{\circ} \}$; an example of computing the wave speed by this slope fitting is shown in the Hovm{\"o}ller diagrams of Section \ref{subsec:hovmuller}. For a perfectly nondispersive wave, this peak-tracking method will correctly identify the phase speed. However, the numerical diffusion and dispersion present in dynamical cores will impact the numerical wave propagation, and thus, the estimated speed. In particular, a horizontal broadening of the maximum surface pressure perturbation pulse due to dispersion or diffusion can lead to overestimations of the peak arrival speed. As the shape of the peak is dependent on grid resolution, the horizontal resolution study in Section \ref{subsec:horiz_res} provides further insight into the accuracy of our peak-tracking method.  \par
For accuracy in the speed computation, we use a one-minute temporal resolution, expressed via a base model time step of $\Delta t_{\text{base}} = 1$ minute. This is a relatively short time step --- the default base time step in CAM for these three models at the approximately $1^{\circ}$ resolution is 30 minutes, with additional temporal subcycling invoked for all dynamical cores. With $\Delta t_{\text{base}} = 1$ minute, we turn off any subcycling for the dynamics (dyn), i.e. $\Delta t_{\text{dyn}}=\Delta t_{\text{base}}$, although we explore subcycling sensitivity in Section \ref{subsec:subcycle}. \par
The simulations are run for 18 hours (1080 minutes), which is sufficient time for the Lamb wave to reach the antipode, ($\lambda_{\text{a}},\phi_{\text{a}}) = ( 90^{\circ}, 0^{\circ})$, with the default parameters. The default vertical grid has a reference model top height of $z_{\text{T}} = 50$ km (the impact of model top height is explored in Section \ref{subsec:zT_study}) and a reference vertical grid spacing of $\Delta z = 500$ m. The hybrid pressure coordinates used in SE and FV3 are generated using the procedure outlined in \textcite{andrews2026mountain}, with the hybrid coefficients computed as per \textcite{laprise1990spectral}. The hybrid coefficients are computed using $T_0=275$ K, which means that using these coefficients in a colder atmosphere slightly reduces $\Delta z$ and $z_{\text{T}}$, and vice versa for a warmer atmosphere. Additionally, the isobaric lid leads to a slightly higher model top in the vicinity of the initial surface pressure disturbance. These differences do not impact our conclusions.

\section{Validation of the Lamb wave}
\label{sec:validation}
We now show the application of the Lamb wave test case in the SE, FV3, and MPAS dynamical cores. In this section, we check that the leading wave excited in this test is indeed the numerical Lamb wave by examining its speed and vertical structure. \par

\subsection{Temperature dependence}
\begin{figure}[htpb]
    \centering
    \includegraphics[width=0.6\linewidth]{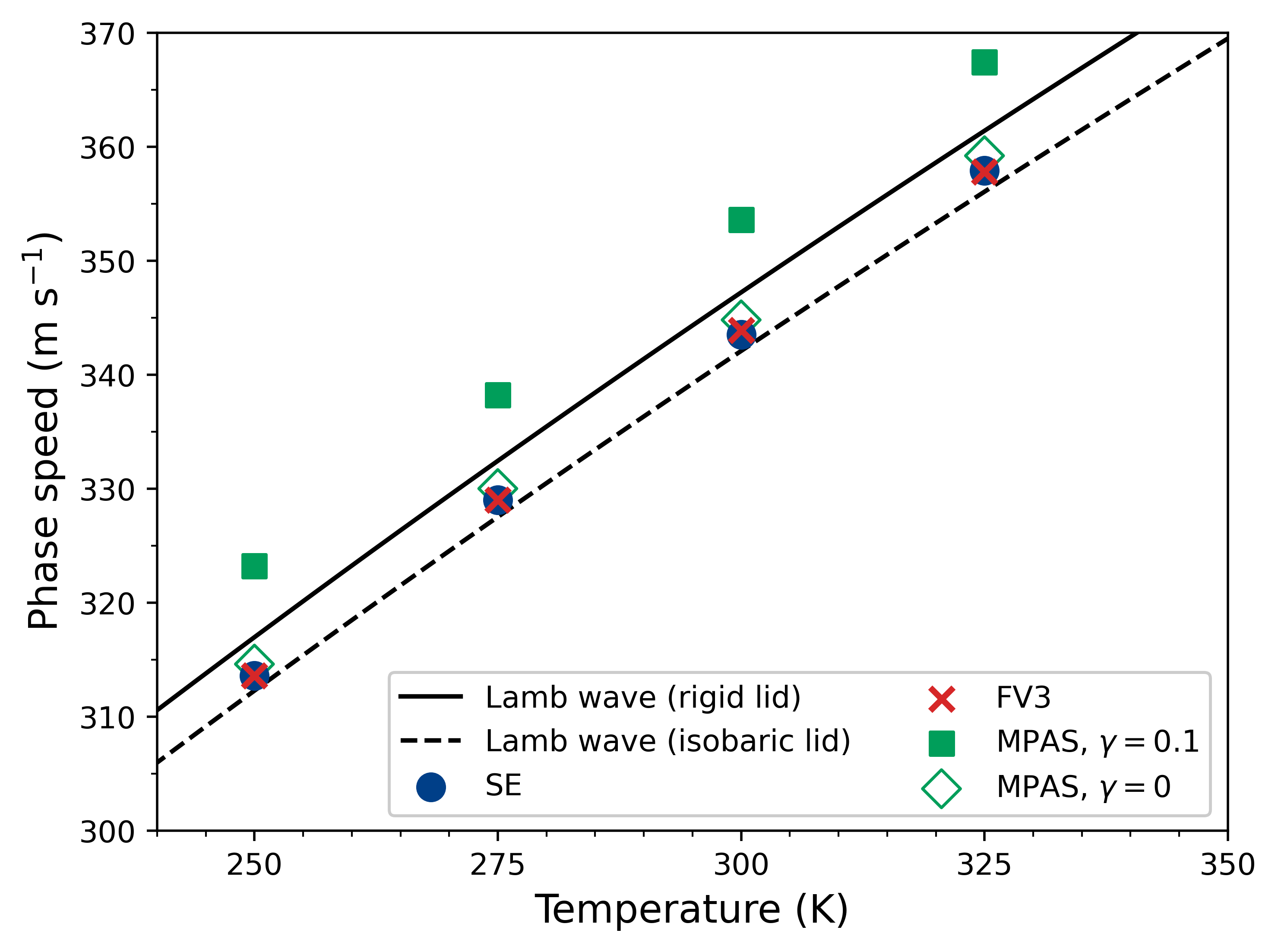}
    \caption{Numerical Lamb wave speeds for the three dynamical cores at the approximate horizontal resolution $1^\circ$ with $z_{\text{T}}$ = 50 km for different isothermal temperatures $T_0$. The theoretical Lamb wave speeds are indicated by the solid and dashed lines for models with rigid and isobaric lids, respectively.}
    \label{fig:p_pert_speeds_with_T0}
\end{figure}

First, we examine the speed of the leading wave and its dependence on the isothermal temperature through simulations with $T_0 \in \{250,275,300,325\}$ K. Figure \ref{fig:p_pert_speeds_with_T0} compares wave speeds from the models to that of the classical Lamb wave (equal to the rigid lid wave speed), which is a function of $\sqrt{T_0}$ (\ref{eq:lamb_speed}), as well as the isobaric lid Lamb wave. The numerical Lamb wave speeds for the dynamical cores are close to the theoretical values (isobaric lid speeds for SE and FV3, and rigid lid speeds for MPAS), and exhibit the expected dependence on temperature. MPAS with the time-adjusted acoustic filter ($\gamma=0.1$) leads to an overestimation of the wave speed, which is a consequence of the filter damping and broadening the peak surface pressure pulse, used for the speed estimates. Deactivation of MPAS' acoustic filter ($\gamma=0$) leads to a wave speed that is slightly slower than predicted. SE and FV3, for a dynamics time step of 1 minute, have a slight overestimation of the wave speed at the $1^{\circ}$ resolution. Impacts of the filters, horizontal resolution, and dynamics time step on the numerical Lamb wave speeds are further explained in Section \ref{sec:sensitivites}. \par
We note that the same hybrid vertical coordinates, generated with $T_0=275$ K, are used for SE and FV3 for all temperature configurations. This means that the model top height varies with $T_0$, and the expected isobaric lid speeds (dashed line in Figure \ref{fig:p_pert_speeds_with_T0}) account for this. MPAS, with its height coordinate, has a fixed 50 km model top for all runs.

\subsection{Model top height}
\label{subsec:zT_study}
\begin{figure}[htpb]
    \centering
    \includegraphics[width=0.6\linewidth]{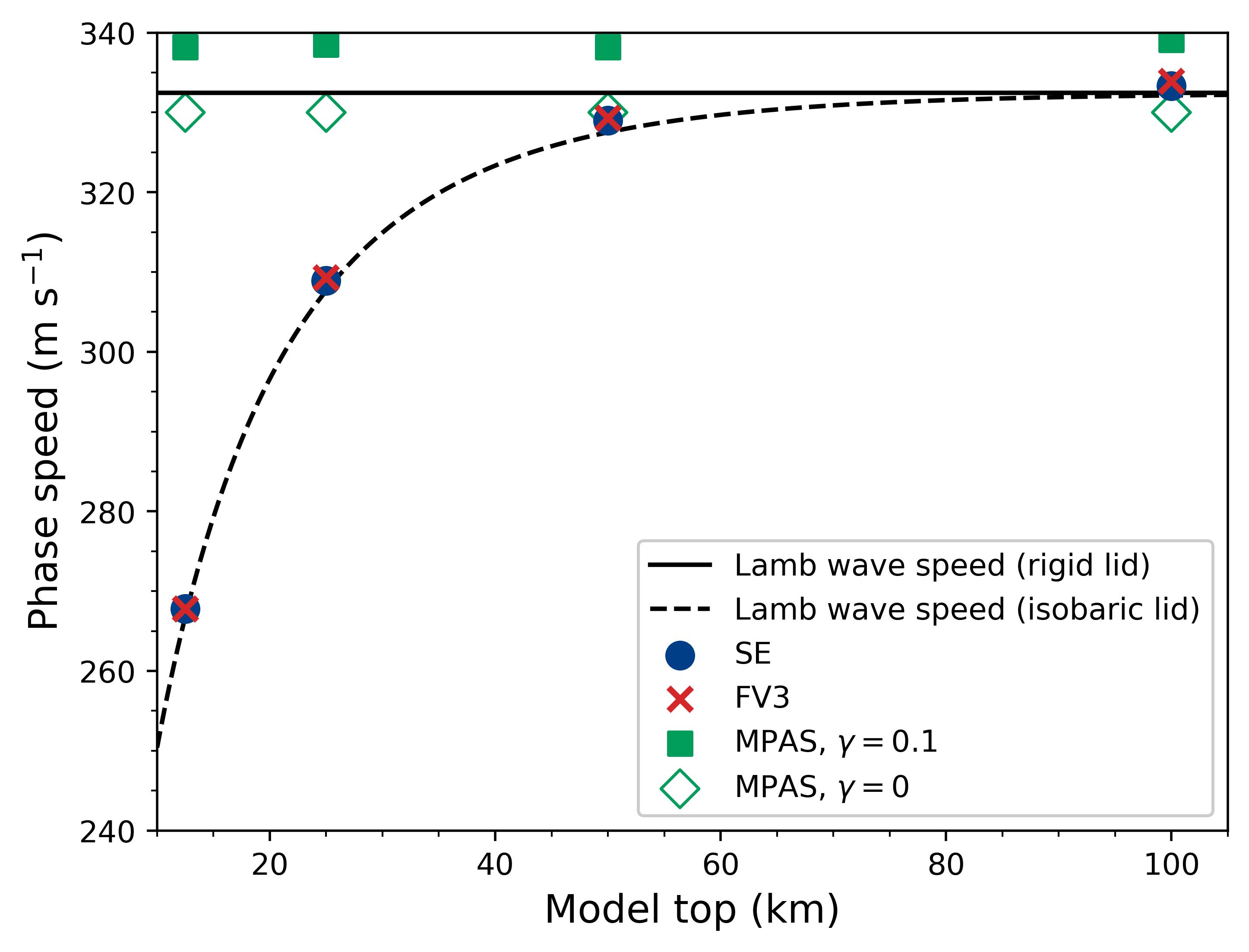}
    \caption{Numerical Lamb wave speeds for different model top heights $z_{\text{T}}$ for the three dynamical cores at the approximately $1^\circ$ horizontal resolution and $T_0$ = 275 K. The classical Lamb wave speed (solid black line) for this temperature is 332.4 m s$^{-1}$.}
    \label{fig:p_pert_speeds_with_zT}
\end{figure}

We next examine the impact of model top height on the speed of the leading wave in Figure \ref{fig:p_pert_speeds_with_zT} by comparing the approximate $1^\circ$ simulations with $z_{\text{T}} \in \{12.5, 25, 50, 100\}$ km in each model, whilst keeping a fixed vertical grid spacing of $\Delta z = 500$ m and reference temperature of $T_0$ = 275 K. As expected from the vertical structure equation theory (Section \ref{subsec:eigenstuff}), the Lamb wave speeds for SE and FV3 vary with $z_{\text{T}}$ as a consequence of the isobaric lid upper boundary condition used with their hybrid pressure coordinate. When using higher model tops, the SE and FV3 Lamb wave speeds asymptote towards the classical Lamb wave speed as denoted by the solid black line. MPAS, with its height coordinate and a rigid lid, has wave speeds that are virtually unchanged with different $z_{\text{T}}$. We again have faster wave speeds than expected in MPAS when using its default time-adjusted acoustic filter with strength $\gamma=0.1$, and slightly slower speeds without the acoustic filter ($\gamma=0$).

\subsection{Vertical profile}
\label{subsec:vert_profile}
\begin{figure}[htpb]
    \centering
    \includegraphics[width=0.8\linewidth]{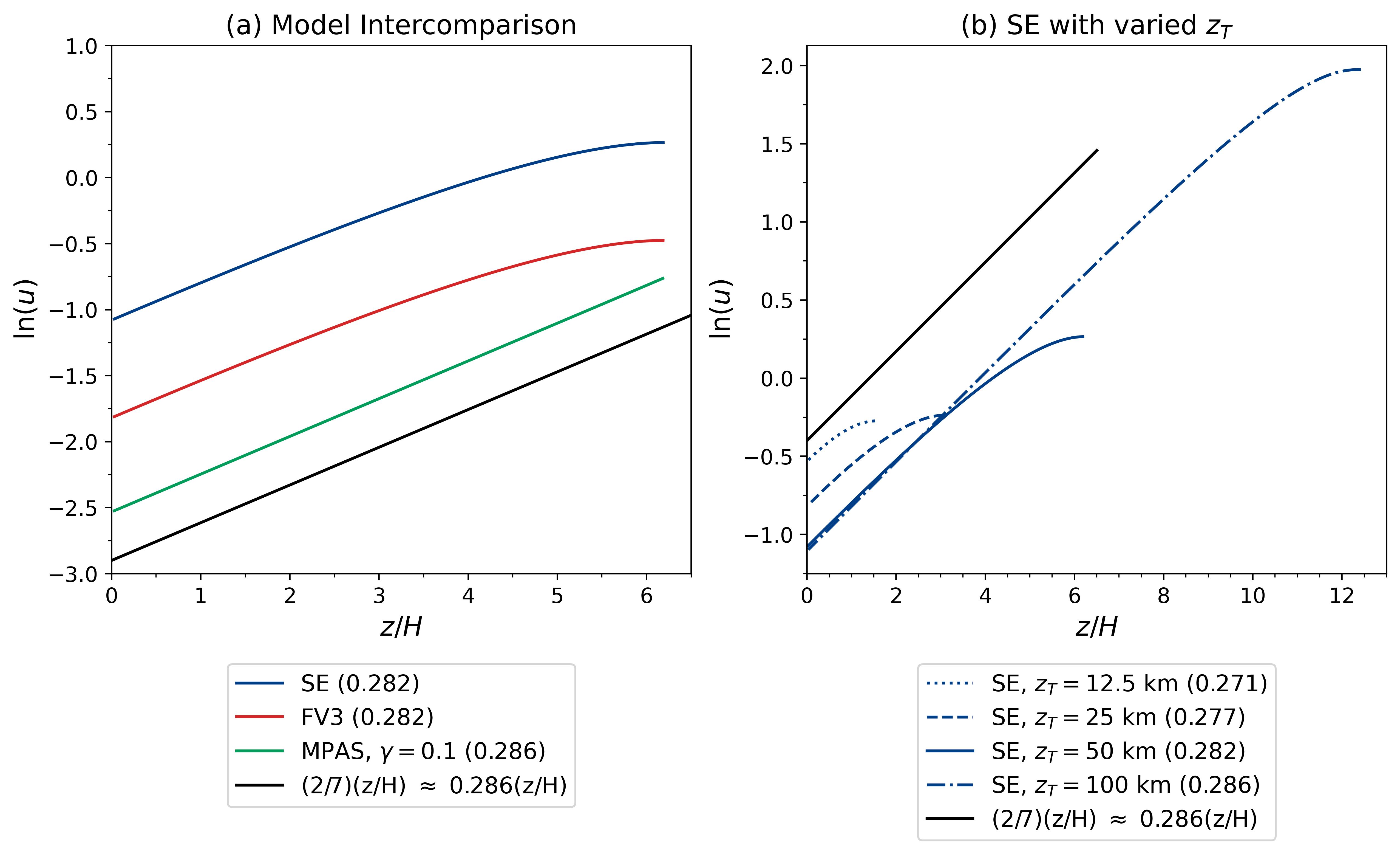}
    \caption{Vertical profile of the zonal velocity of the numerical Lamb wave, after six hours. The slope of $\ln(u) $ with respect to $z/H$ is estimated from linear regression over the first scale height $H$ and quoted in the legends. Note, we take $u$ here to be normalised by $1 ~\text{m} ~\text{s}^{-1}$ such that $\ln(u)$ is dimensionless. Classical Lamb wave theory predicts slopes of $2/7 \approx 0.286$ for $\ln(u)$ (black lines). The three dynamical cores are compared with $z_{\text{T}}=50$ km, $T_0$ = 275 K, and the approximate $1^\circ$ horizontal resolution in (a), whilst (b) compares different model top heights in SE. The slopes for the 12.5 km and 25 km model tops use the first 0.25 and 0.5 scale heights, respectively, to avoid incorporating the slope tapering near the model top in isobaric lid models.}
    \label{fig:vert_profile_p_pert}
\end{figure}

We now examine the vertical structure of the leading wave. The external Lamb wave should have zonal (and meridional) velocities that increase exponentially with height as $\ln(u) \propto 2z/(7H) \approx 0.286 z/H$ (\ref{eq:lamb_u_vert_dependence}); this is in contrast to the $\exp(z/(2H))$ amplitude modulation of the sinusoidal profile for gravity waves (\ref{eq:grav_u_vert_dependence}). Figure \ref{fig:vert_profile_p_pert}a shows that the vertical exponential growth is very close to that expected from classical Lamb wave theory for all three dynamical cores, over the majority of the domain. \par
MPAS, with its rigid lid, maintains the expected 2/7 exponential vertical profile right up to the model top (\ref{eq:L0_vert_profile}). However, for the isobaric lid models of SE and FV3, there is a tapering of the slope such that the vertical gradient of the zonal velocity approaches zero at the model top. This is explained by the proportionality between the vertical structure function and the zonal wind gradient: $L_n(z) \propto du/dz$ (this relationship is derived in Appendix \ref{append:Ln_to_dUdz}). Hence, the fastest mode in an isobaric lid model has the classical Lamb wave structure for the majority of the domain, but deviates at the highest model levels to obey the upper boundary condition of $L_n(z)=0$. Figure \ref{fig:vert_profile_p_pert}a also shows that SE exhibits the strongest Lamb wave amplitudes, followed by FV3, then MPAS with its default time-adjusted acoustic filter ($\gamma = 0.1$). The amplitude variations between models are further discussed in Section \ref{subsec:amplitude_comps}. \par
Figure \ref{fig:vert_profile_p_pert}b compares vertical profiles in SE over a range of model top heights. Each value of $z_{\text{T}}$ leads to the expected 2/7 vertical dependence away from the model top. There is increasing agreement with higher $z_{\text{T}}$, which reflects the isobaric lid Lamb wave tending to the classical Lamb wave in the limit of $z_{\text{T}} \rightarrow \infty$. Obtaining the 2/7 slope for the low model top of 12.5 km, albeit for a smaller proportion of the domain than with a larger $z_{\text{T}}$, shows that the vertical characteristics of this wave are preserved between the internal ($z_{\text{T}} < 8H/3$) and external ($z_{\text{T}} > 8H/3$) regimes of the numerical Lamb wave (for this test case, the critical height where these regimes change is $z_{\text{T}} = 8H/3 \approx 21.5$ km). Hence, although the change in sign of $(m^*)^2$ means that the $z_{\text{T}} = 12.5$ km wave is an $n=0$ baroclinic mode and the others ($z_{\text{T}} \in \{ 25, 50, 100\}$ km) are barotropic modes according to the classification of \textcite{de2020barotropic}, we identify all of these as the numerical Lamb wave for an isobaric lid, i.e. the discretisation of the classical Lamb wave in a model with a fixed pressure upper boundary condition. \par
For a higher model top, the amplitude of the Lamb wave is reduced at a given altitude (Figure \ref{fig:vert_profile_p_pert}b). This may reflect the energy from the initial condition being distributed over a larger domain. If we consider the Lamb wave kinetic energy density, which decreases with height proportionally to $\exp(-3z/(7H))$ (Section \ref{subsec:govern_equations}), its value above 44 km is less than 10\% of that at the surface; this may explain the smaller difference between the amplitudes with a 50 km or 100 km model top, compared to 12.5 km or 25 km model tops. \par

\begin{figure}[htpb]
    \centering
    \includegraphics[width=\linewidth]{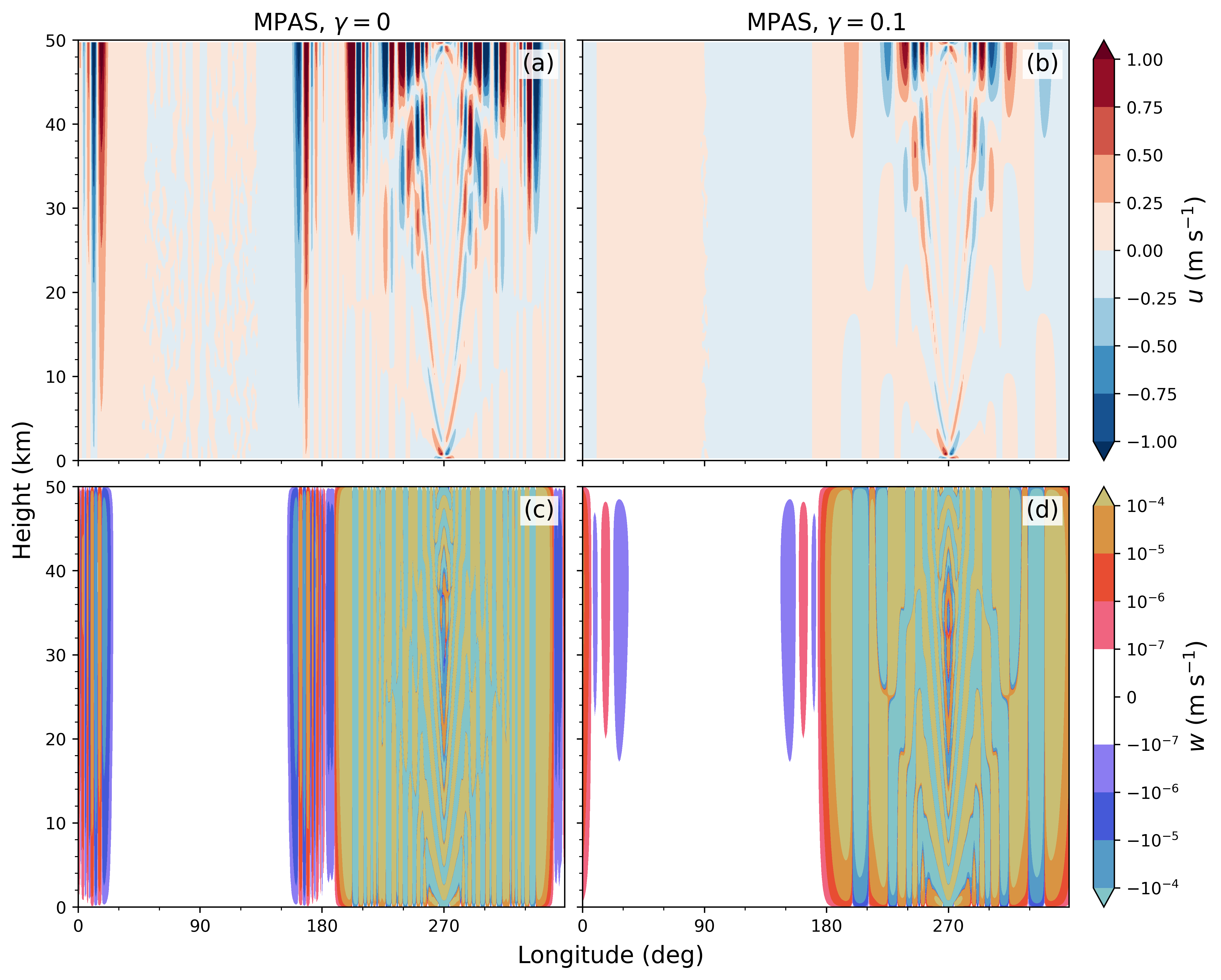}
    \caption{Longitude-height cross-sections along the equator of the (a,b) zonal velocity and (c,d) vertical velocity after ten hours in MPAS with $z_{\text{T}}=50$ km, $T_0$ = 275 K, and the approximate horizontal resolution $1^\circ$. The default time-adjusted acoustic filter is turned off in (a,c) with $\gamma=0$ and is present in (b,d) with strength $\gamma=0.1$. The colour bar is saturated for the trailing gravity waves to highlight the position, structure and amplitude of the leading Lamb wave.}
    \label{fig:MPAS_u_w_cross_section}
\end{figure}

The vertical structure of the Lamb wave is further examined in Figure \ref{fig:MPAS_u_w_cross_section}, with vertical cross-sections of the zonal and vertical velocities along the equator for MPAS. As MPAS is a nonhydrostatic model, $w$ is a prognostic variable, unlike in hydrostatic dynamical cores, where it is diagnosed. The cross-sections show the fields after ten hours, when the rigid lid Lamb wave fronts are at approximately $\lambda=162^{\circ}$ and $\lambda=18^{\circ}$. Without the time-adjusted acoustic filter ($\gamma=0$), the Lamb wave can be clearly identified in the zonal velocity field through its exponentially increasing profile with height (Figure \ref{fig:MPAS_u_w_cross_section}a), whilst the filtered simulation ($\gamma=0.1$) has a significantly weaker Lamb wave (Figure \ref{fig:MPAS_u_w_cross_section}b). The $w$ cross-sections show that the numerical Lamb wave exhibits a small vertical velocity that is close to the $w=0$ condition for classical Lamb waves: without the filter ($\gamma=0$) $|w|<10^{-4} ~\text{m} ~\text{s}^{-1}$ (Figure \ref{fig:MPAS_u_w_cross_section}c) whilst $|w|<10^{-6} ~\text{m} ~\text{s}^{-1}$ (Figure \ref{fig:MPAS_u_w_cross_section}d) for $\gamma=0.1$. The cross-sections also highlight the generation of the trailing gravity waves. These gravity wave speeds reduce with larger vertical mode number $n$, where $n$ can be observed through the number of sign changes in the vertical profiles.

\section{Dynamical core sensitivities}
\label{sec:sensitivites}
We now compare the numerical Lamb wave between the models, as well as examine the impact of model parameters such as the horizontal resolution, MPAS' time-adjusted acoustic filter, and the dynamics time step size. Hovm\"{o}ller and wavenumber-frequency diagrams are also provided for alternative interpretations of the numerical Lamb waves. Lastly, parameters that had no impact on the numerical Lamb wave are briefly noted.

\subsection{Amplitude comparison}
\label{subsec:amplitude_comps}
\begin{figure}[htpb]
    \centering
    \includegraphics[width=0.5\linewidth]{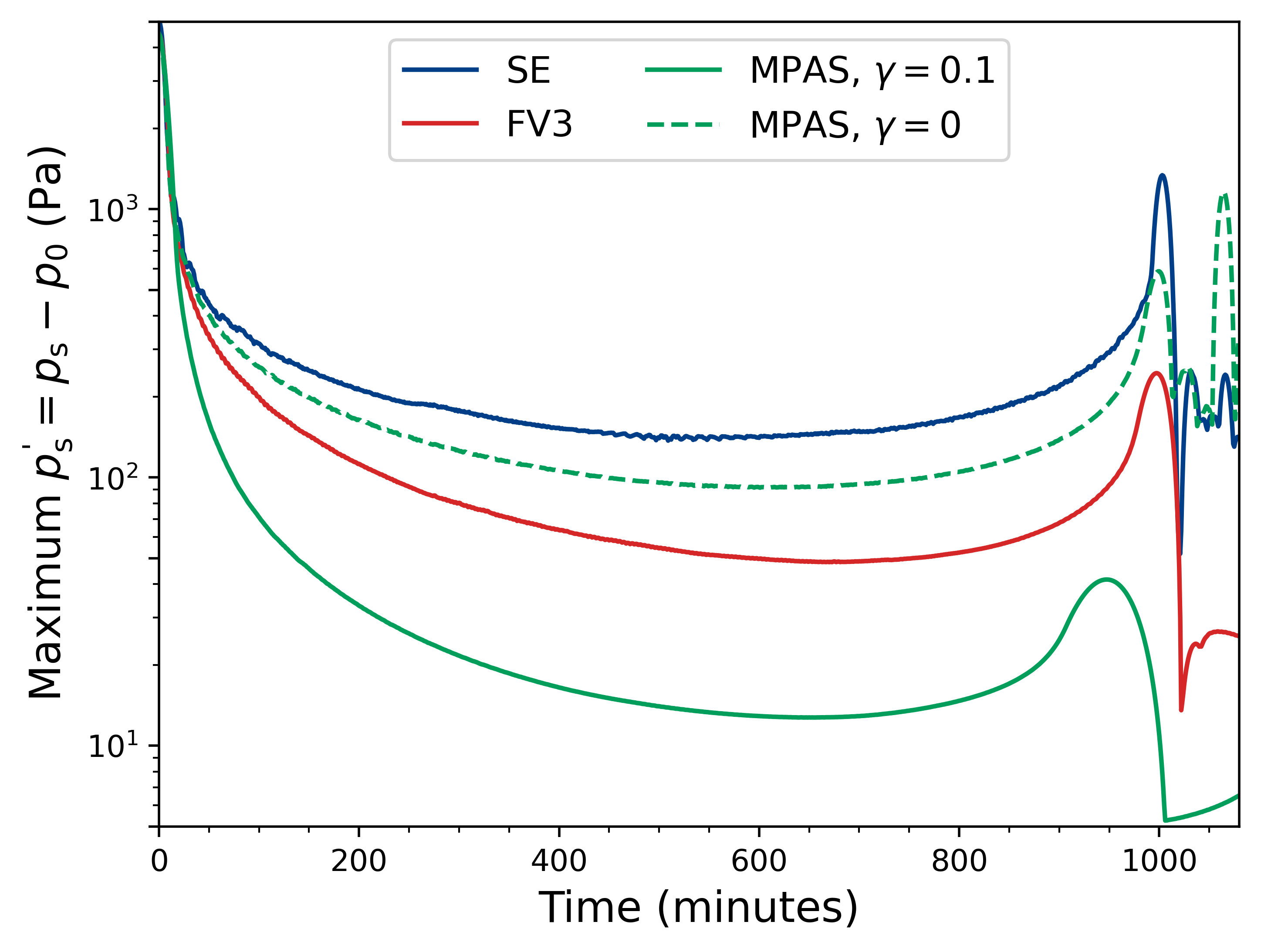}
    \caption{Time series of the maximum surface pressure perturbation in the Lamb wave test case with $z_{\text{T}}=50$ km, $T_0$ = 275 K, and the approximate horizontal resolution $1^\circ$.}
    \label{fig:p_pert_ps_max_time_series}
\end{figure}

In the vertical profile comparisons (Figure \ref{fig:vert_profile_p_pert}a), the amplitude of the zonal wind associated with the numerical Lamb wave was shown to vary between dynamical cores. This amplitude difference is further examined through time series of the maximum of the surface pressure perturbation field, $p_{\text{s}}' = p_{\text{s}}-p_0$, which identifies the numerical Lamb wave (Figure \ref{fig:p_pert_ps_max_time_series}). For the first ten hours, the $p_{\text{s}}'$ maximum reduces from its initial value of 50 hPa, before growing around 900--1000 minutes as the wave approaches the antipode. The Lamb wave in SE has the largest amplitude for all times up to reaching the antipode. The amplitude in MPAS is considerably weaker than the other models with its default time-adjusted acoustic filter ($\gamma=0.1$), whilst switching off the filter ($\gamma=0$) increases the Lamb wave strength by an order of magnitude, making it stronger than the FV3 Lamb wave. Hence, MPAS' time-adjusted acoustic filter successfully damps the Lamb wave in this test. However, as a consequence the damped wave signal spreads out and causes the $p_{\text{s}}'$ maximum to reach the antipode faster. This apparent ``supersonic'' numerical Lamb wave for MPAS with $\gamma=0.1$, as was seen for different isothermal temperatures (Figure \ref{fig:p_pert_speeds_with_T0}) or model top heights (Figure \ref{fig:p_pert_speeds_with_zT}), can again be attributed to the acoustic filter broadening the Lamb wave $p_{\text{s}}'$ signal and impacting the speed estimate (Section \ref{subsec:sim_considers}). \par

\subsection{Temporal subcycling}
\label{subsec:subcycle}
As discussed in Section \ref{subsec:sim_considers}, the small base time step of $\Delta t_{\text{base}}=60$ s means that a subcycling of the base time step to shorten the dynamics time step is not required for stability at the approximately $1^{\circ}$ resolution, but it can be optionally used, as we explore here. For SE and MPAS without the time-adjusted acoustic filter ($\gamma=0$), subcycling has minimal impact on the numerical Lamb wave. MPAS with the time-adjusted acoustic filter ($\gamma=0.1$) exhibits a small sensitivity, with subcycling further reducing the already small Lamb wave amplitude (not shown). For FV3, there is a strong sensitivity to subcycling, which we now discuss. \par
For FV3, there are two pertinent subcycling parameters. The first is $k_{\text{split}}$, which determines the number of vertical remappings of the floating Lagrangian levels within each base time step. The second is $n_{\text{split}}$, which determines the number of dynamics substeps between each Lagrangian remapping. The dynamics time step is thus
\begin{equation}
    \Delta t_{\text{dyn}} = \frac{\Delta t_{\text{base}}}{k_{\text{split}} \times n_{\text{split}}}.
\label{eq:subcycle}
\end{equation}

\noindent By default, FV3 uses $k_{\text{split}}=2, n_{\text{split}}=6$ for a base time step of $\Delta t_{\text{base}} = 1800$ s at the C96 ($\approx 1^\circ$) horizontal resolution. In the FV3 results discussed so far with the default time step for the test case of $\Delta t_{\text{base}} = 60$ s, no subcycling was selected via $k_{\text{split}}=1$ and $n_{\text{split}}=1$, leading to $\Delta t_{\text{dyn}} = 60$ s.\par
\begin{figure}[htpb]
    \centering
    \includegraphics[width=0.7\linewidth]{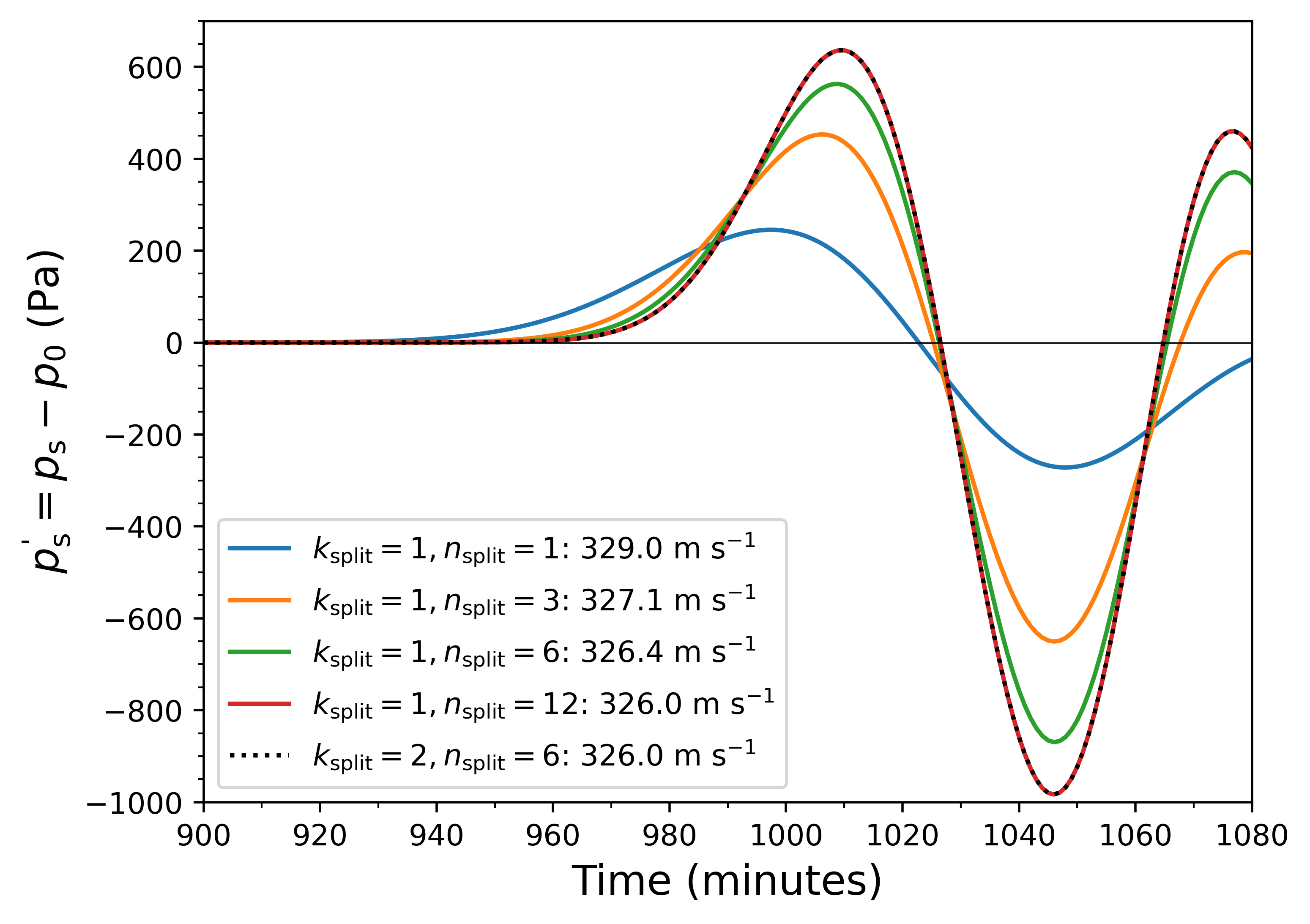}
    \caption{Time series of the surface pressure perturbation at the antipode, $(\lambda_{\text{a}}, \phi_{\text{a}})=(90^{\circ}, 0^{\circ})$, for FV3 C96 ($\approx 1^\circ$ horizontal resolution) with different temporal subcycling configurations that correspond to the dynamics time steps 60 s (blue), 20 s (orange), 10 s (green), and 5 s (red and dotted black), respectively. The legend shows the numerical Lamb wave speeds for these configurations with an isobaric lid, $z_{\text{T}} = 50$ km, and $T_0 = 275$ K. The expected Lamb wave speed from theory is 327.5 m s$^{-1}$.}
    \label{fig:fv3_subcycling}
\end{figure}

Figure \ref{fig:fv3_subcycling} compares a selection of FV3 subcycling configurations for $\Delta t_{\text{base}} = 60$ s. The figure shows that a smaller dynamics time step from increased subcycling leads to a larger numerical Lamb wave amplitude and a reduced speed. The sensitivity is only to the dynamics time step size and not the frequency of vertical remapping; this is seen through the alignment of the $k_{\text{split}}=2, n_{\text{split}}=6$ and $k_{\text{split}}=1, n_{\text{split}}=12$ configurations, which both have twelve dynamics subcycles per base time step and $\Delta t_{\text{dyn}}=5$ s. Hence, our discussion from here on only considers the number of dynamics subcycles per base time step ($k_{\text{split}} \times n_{\text{split}}$). \par
FV3's sensitivity to subcycling of the dynamics can be explained from the analysis of \textcite{skamarock2008linear}. FV3's forward-backward time stepping method on the CD-grid is first-order accurate and contains a diffusion term that damps horizontally divergent motions such as the Lamb wave. Increased subcycling leads to less diffusion of the divergent motions and a sharper representation of the Lamb wave propagation. In the next section, we examine how the number of dynamics subcycles interacts with finer horizontal resolutions.

\subsection{Horizontal resolution}
\label{subsec:horiz_res}
We study the impact of horizontal resolution by running each model at approximately $0.5^{\circ}$ (50--60 km grid spacings) and $0.25^{\circ}$ (25--30 km grid spacings) horizontal resolutions, in addition to the default $1^{\circ}$ case. For MPAS, the $0.25^{\circ}$ resolution requires subcycling such that the acoustic time step is half the base time step of $\Delta t_{\text{base}}=60$ s. FV3 is also unstable without subcycling at the finest resolution, so $\Delta t_{\text{dyn}}=60$ s is not shown for the $0.25^{\circ}$ resolution. \par
\begin{figure}[htpb]
    \centering
    \includegraphics[width=0.9\linewidth]{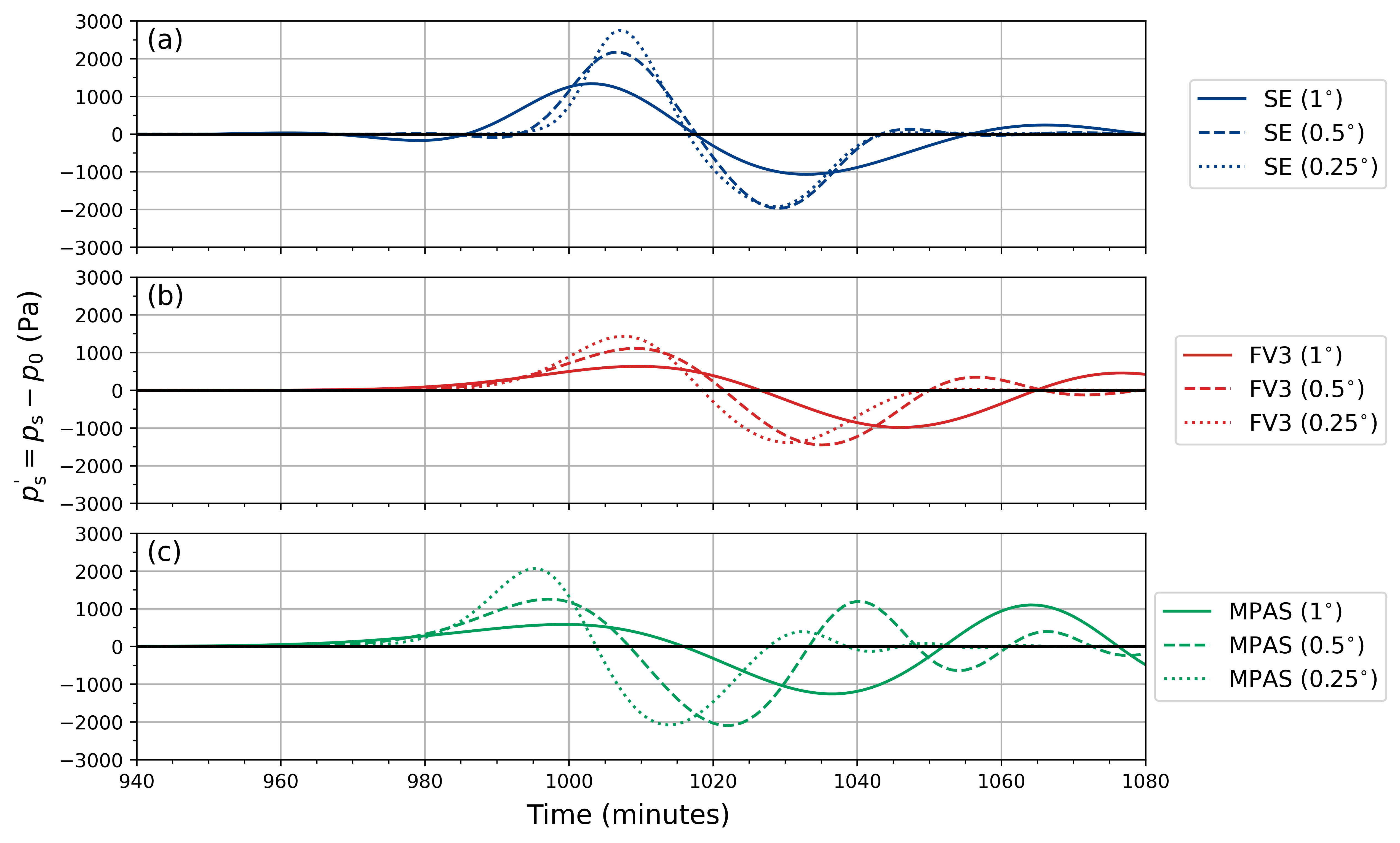}
    \caption{Time series of the surface pressure perturbation at the antipode, at $\approx 1^{\circ}$ (solid), $\approx  0.5^{\circ}$ (dashed) and $\approx 0.25^\circ$ (dotted) horizontal resolutions for (a) SE, (b) FV3, and (c) MPAS, all with $z_{\text{T}} = 50$ km and $T_0 = 275$ K. MPAS is shown without the time-adjusted acoustic filter ($\gamma=0$) and all resolutions for FV3 use twelve dynamics subcycles ($\Delta t_{\text{dyn}}=5$ s).}
    \label{fig:horiz_res_study}
\end{figure}

Figure \ref{fig:horiz_res_study} shows time series of the surface pressure perturbation at the antipode, across the horizontal resolutions. FV3 is shown with 12 subcycles of the dynamics, so $\Delta t_{\text{dyn}}=5$ s. For SE and MPAS (shown without the time-adjusted acoustic filter), increasing the horizontal resolution leads to a larger numerical Lamb wave amplitude and reduced dispersion. Increasing the horizontal resolution from $1^{\circ}$ slows the numerical Lamb wave in SE, but leads to faster speeds for FV3 (with 12 dynamics subcycles) and MPAS (with $\gamma=0$). \par
\begin{figure}
    \centering
    \includegraphics[width=0.7\linewidth]{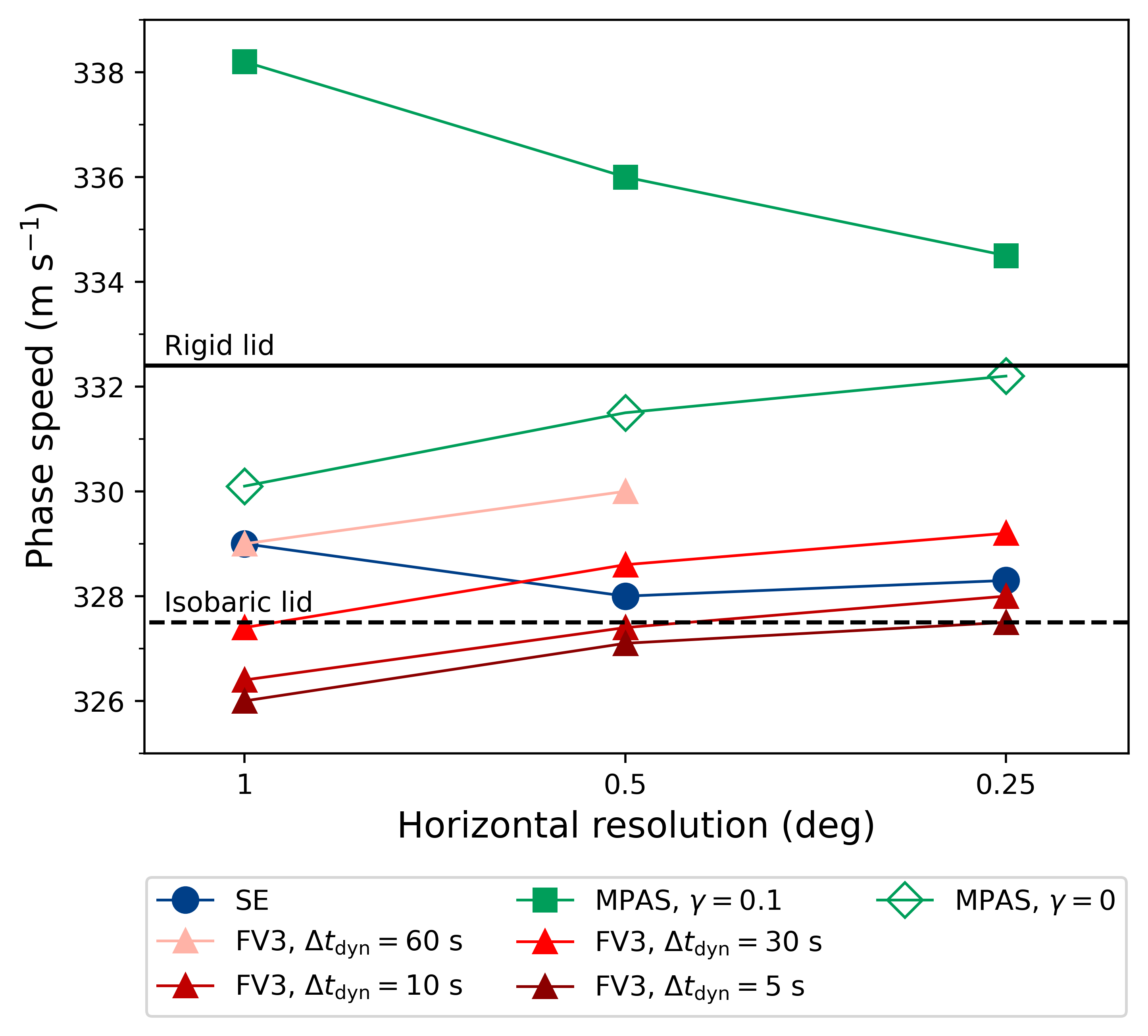}
    \caption{Numerical Lamb wave speeds as a function of horizontal resolution for the three dynamical cores with $z_{\text{T}} = 50$ km and $T_0 = 275$ K. The horizontal lines indicate the theoretical wave speeds with a rigid (solid) or isobaric (dashed) lid. MPAS is compared with and without the time-adjusted acoustic filter. FV3 is compared with four dynamics time step sizes, with $\Delta t_{\text{dyn}}=60$ s and $0.25^{\circ}$ not shown as this is unstable.}
    \label{fig:horiz_res_scatter}
\end{figure}

Figure \ref{fig:horiz_res_scatter} further explores the impact of horizontal resolution, by plotting the numerical Lamb wave speed as a function of horizontal resolution. For SE, the $1^{\circ}$ run is faster than expected, with $0.5^{\circ}$ and $0.25^{\circ}$ being close to the theoretical speed with an isobaric lid. For MPAS, simulations with and without the time-adjusted acoustic filter converge to the rigid lid (classical) Lamb wave speed with increased resolution. However, the theoretical speed is approached from above with the time-adjusted acoustic filter ($\gamma=0.1$), and from below without the filter ($\gamma=0$). For FV3, we compare configurations with 2, 6, and 12 subcycles for all resolutions, which correspond to a dynamics time step (\ref{eq:subcycle}) of $\Delta t_{\text{dyn}}=30, 10,$ and 5 s, respectively. The use of no subcycling and $\Delta t_{\text{dyn}}=60$ s is only shown for the stable $1^{\circ}$ and $0.5^{\circ}$ resolutions. For each subcycling configuration, the numerical Lamb wave speed increases with a finer horizontal resolution. Reduced subcycling (larger $\Delta t_{\text{dyn}}$) increases the speed at each resolution, as a result of increased diffusion of the divergent motions acting on the surface pressure peak (Figure \ref{fig:fv3_subcycling}). For the FV3 CAM default of 12 dynamics subcycles ($\Delta t_{\text{dyn}}=5$ s), there is convergence to the expected speed with increasing horizontal resolution, whilst fewer subcycles lead to faster than expected speeds at $0.25^{\circ}$. \par

\FloatBarrier

\subsection{Hovm\"{o}ller and zonal wavenumber-frequency diagrams}
\label{subsec:hovmuller}
\begin{figure}
    \centering
    \includegraphics[width=\linewidth]{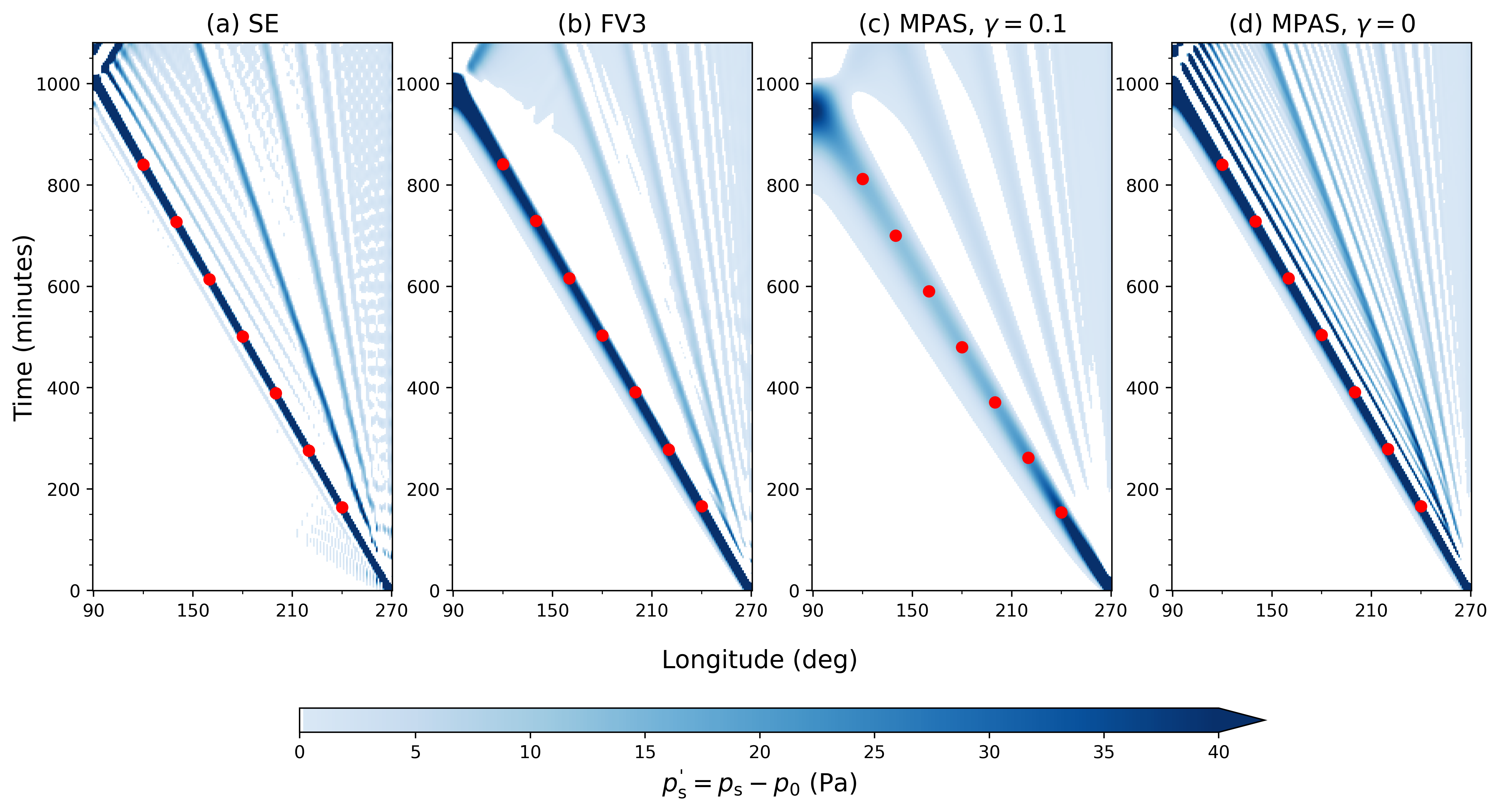}
    \caption{Hovm\"{o}ller diagrams of the surface pressure perturbation $p_{\text{s}}^{'}$ in each $1^\circ$ model with $z_{\text{T}} = 50$ km and $T_0 = 275$ K, for the longitude range $\lambda \in [\lambda_a, \lambda_c] = [90^{\circ}, 270^{\circ}]$. Small perturbations of $p_{\text{s}}^{'} < 0.1$ Pa are shown in white. The colour bar is saturated to show the presence of gravity waves along with the numerical Lamb wave. The red dots correspond to the times and locations used to estimate the wave speeds, as described in Section \ref{subsec:sim_considers}.}
    \label{fig:hovmoller}
\end{figure}

We now provide some alternative methods for analysing the numerical Lamb waves from the dynamical cores: Hovm\"{o}ller and zonal wavenumber-frequency diagrams. 
Hovm\"{o}ller diagrams (Figure \ref{fig:hovmoller}) show the spatiotemporal evolution of the surface pressure perturbation $p_{\text{s}}'$. The numerical Lamb wave in each simulation is identified by the strongest signal, which reaches the antipode ($\lambda_{\text{a}}=90^{\circ}$) after about 1000 minutes. Figure \ref{fig:hovmoller} also shows the times that the $p_{\text{s}}'$ maximum reaches the selected longitudinal locations used to compute the Lamb wave speed (Section \ref{subsec:sim_considers}). \par
For SE (Figure \ref{fig:hovmoller}a), the Lamb wave shows minimal dispersion (a small longitudinal range at each time), whereas there is more spread for FV3 (Figure \ref{fig:hovmoller}b) as a consequence of no subcycling, which leads to more diffusion of the Lamb wave (Section \ref{subsec:subcycle}). MPAS (Figures \ref{fig:hovmoller}c,d) with the time-adjusted acoustic filter ($\gamma=0.1$) has a much weaker and more diffused signal than the other Hovm\"{o}ller diagrams, again showing the significant impact of this filter compared to the unfiltered case ($\gamma=0$). The Hovm\"{o}ller diagrams also show the presence of gravity waves at different mode numbers $n$, which travel at slower speeds and do not reach the antipode within 18 hours (1080 minutes).  \par

\begin{figure}
    \centering
    \includegraphics[width=\linewidth]{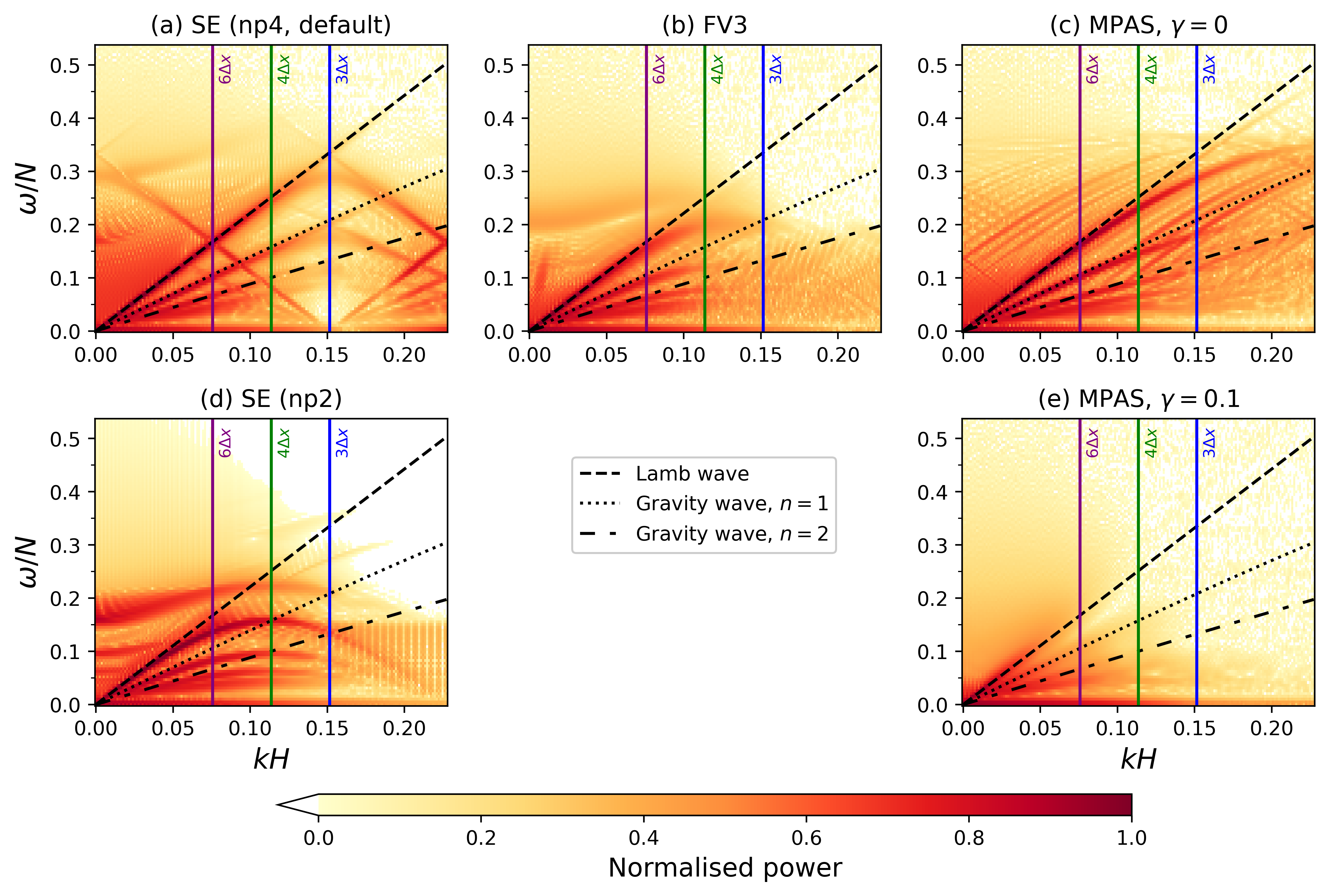}
    \caption{Zonal wavenumber-frequency plots of the equatorial surface pressure power spectra (the squared Fourier representation of $p_{\text{s}}'$) for $1^\circ$ simulations, $z_{\text{T}} = 50$ km and $T_0 = 275$ K. Using the approach of \textcite{Vallis}, the wavenumber is normalised by the isothermal scale height $H$, and the frequency by the Brunt-V{\"a}is{\"a}l{\"a} frequency $N$ (\ref{eq:brunt_vais_freq}). The power from each dynamical core is normalised to have a maximum of 1. The default SE configuration with ne30np4 is shown in (a), and with the lower-order ne90np2 configuration in (d). MPAS is shown with and without the time-adjusted acoustic filter in (e) and (c), respectively. Scales of $6\Delta x, 4\Delta x, 3\Delta x$ are indicated by vertical lines (note that the $2\Delta x$ scale represents the largest resolvable wavenumber).}
    \label{fig:omega_k}
\end{figure}

We next provide a spectral perspective through zonal wavenumber-frequency plots of the spatially and temporally Fourier-transformed surface pressure ($\hat{p}_{\text{s}}$) along the equator (Figure \ref{fig:omega_k}). A spectral power of $|\hat{p}_{\text{s}}|^2$ is computed and normalised so that the largest spectral power in each model is unity; this normalisation choice highlights the distribution of power over wavenumbers and frequencies within each model. The zonal wavenumber is denoted as $k$, and the effective grid spacing of the output is $\Delta x = 2\pi a/N_e$, for Earth's radius $a=6371.22$ km and $N_e$ the number of nodes along the equator ($N_e=360$ for $1^{\circ}$). \par
The zonal wavenumber-frequency spectrum for SE (Figure \ref{fig:omega_k}a) shows that the numerical Lamb wave closely follows the expected dispersion relation up to the $4 \Delta x$ scale, with $\omega$ reaching its maximum value at $3 \Delta x$. At smaller scales (larger $k$) than $3\Delta x$, there is a sharp deviation from the theoretical dispersion relation, with the energy following a path of $\partial \omega /\partial k < 0$. This transition at $3 \Delta x$ is also seen for the gravity wave spectra, most clearly for the $n=1$ mode. This result is in agreement with linear dispersion analyses of the spectral element method such as \textcite{melvin2012dispersion}\footnote{\textcite{melvin2012dispersion} considers a one-way wave equation which is typically used to examine the dispersion of gravity waves, but this general analysis also applies to the propagation of the numerical Lamb wave. A comparison of numerical dispersion relations for element orders 1--6 is shown in Figure 11 of \textcite{melvin2012dispersion}.} and \textcite{ullrich2018hyperdiffusion}. Such analyses show that cubic spectral elements, which are the default choice for the SE dynamical core (the np4 configuration), lead to negative $\partial \omega/\partial k$ at scales smaller than $3 \Delta x$. We note that cubic elements generate three branches of numerical dispersion relation. For linear elements (the np2 configuration), where there is only a single numerical dispersion relation branch, $\partial \omega /\partial k < 0$ beyond the $4 \Delta x$ scale, which is the upper half of the wavenumbers. A zonal wavenumber-frequency plot from an np2 simulation with 90 cells per panel edge, which matches the average nodal spacing of np4 at a $1^{\circ}$ resolution with 30 cells per panel edge, generates the bending of the numerical Lamb wave dispersion relation at $4 \Delta x$ (Figure \ref{fig:omega_k}d), consistent with linear dispersion theory. \par
The zonal wavenumber-frequency diagram for FV3 (Figure \ref{fig:omega_k}b) highlights model diffusion, with weak Lamb and gravity wave spectra beyond the $4 \Delta x$ scale. MPAS with $\gamma=0$ has power in the Lamb wave mode up to the $2\Delta x$ grid scale, with deviations from the theoretical Lamb wave dispersion relation occurring between the $6 \Delta x$ and $4 \Delta x$ scales. By contrast, use of the time-adjusted acoustic filter ($\gamma=0.1$) removes power from the small-scale components of the Lamb and gravity waves (Figure \ref{fig:omega_k}e), and leads to the much weaker waves seen in its Hovm\"{o}ller diagram (Figure \ref{fig:hovmoller}c). \par

\FloatBarrier

\subsection{Model insensitivities}
\label{subsec:other_tests}
Finally, we briefly mention some other parameters within the dynamical cores that were determined to have negligible impact on the numerical Lamb wave in our test (no figures shown).

\begin{itemize}
    \item Nonhydrostatic versions of SE and FV3 compared to their hydrostatic versions. The Lamb wave differences between the hydrostatic and nonhydrostatic configurations were very minor.
    \item Vertical resolution. SE simulations with equidistant vertical grid spacings of $\Delta z \in \{ 250, 500, 1000 \}$ m (all with a model top at $z_{\text{T}}=50$ km) revealed that the phase speeds and amplitudes of the Lamb waves were insensitive to these variations of the vertical resolution.
    \item The strength of horizontal divergence damping in FV3. Reducing the dimensionless coefficient for sixth-order horizontal divergence damping from its default strength of 0.15, even down to zero, had no impact. We note that FV3 is typically unstable without horizontal divergence damping, but was stable for 18 hours in this test.
    \item FV3's CAM-default transport configuration compared to a virtually inviscid configuration. CAM-FV3's default advection scheme applies monotonicity constraints, whilst the virtually inviscid choice is almost unlimited \parencite{harris2021scientific_FV3_doc}. Insensitivity to this change indicates that monotonicity constraints on transport are not impactful in this test case with a smooth initial perturbation.
    \item Initial pressure perturbation amplitude. A weaker initial surface pressure perturbation of 1 hPa, compared to the default 50 hPa used here, generated analogous wave dynamics. Simulations in both MPAS and FV3 showed the Lamb wave amplitude reducing by roughly a factor of fifty for the weaker initial perturbation, but with unaffected speed and shape (e.g. for FV3, the ratio of peak antipodal amplitudes is 50.4). This indicates that the selected initial 50 hPa surface pressure perturbation generates an approximately linear dynamic response.
    \item Other default MPAS dissipation settings. The activated damping of vertically propagating acoustic waves via an implicit vertical off-centring, the second-order horizontal Smagorinsky diffusion, or the fourth-order horizontal filter via hyperdiffusion had minimal impact on the Lamb wave signal over the first 18 hours. The only notable sensitivity to the dissipation mechanisms was caused by MPAS' horizontal acoustic filter, as discussed earlier.
    \item Variations of MPAS' three-stage Runge-Kutta (RK) time stepping technique. MPAS' RK algorithm is second-order accurate for nonlinear motions and can be configured as a second-order or third-order accurate time stepping scheme for linear motions. Testing these two RK variants in MPAS without the acoustic filter led to almost indistinguishable Lamb wave results with maximum surface pressure differences around 0.015 Pa after 18 hours.
\end{itemize}

\FloatBarrier

\section{Conclusions and Discussion}
\label{sec:conclusions}
This work examined numerical Lamb waves, which are the representation of the Lamb wave mode in a discrete model such as a dynamical core. We presented a simple test case in an initially isothermal atmosphere at rest, which excites a dominant Lamb wave along with weaker trailing gravity waves. We validated that three dynamical cores (SE, FV3, and MPAS) generated numerical Lamb waves, as determined from speeds that were sensitive to the initial isothermal temperature, and vertical profiles of the horizontal velocities that were proportional to $\exp(2z/(7H))$. As the dynamics of this test are linear, it enables the examination of wave dispersion properties through wavenumber-frequency spectra. Deviations from the theoretical Lamb wave dispersion relation in the SE spectra were consistent with analyses of the spectral element method, with $\partial \omega/\partial k < 0$ beyond scales of $4 \Delta x$ and $3 \Delta x$ for linear and cubic elements, respectively. \par
The choice of the vertical coordinate and the corresponding model boundary conditions have a considerable impact on the numerical Lamb waves by determining the supported eigenmodes, identified through the vertical structure equation. A rigid lid with a vanishing vertical velocity $w$ at the upper boundary leads to numerical Lamb waves that are insensitive to the model top height, as shown in the MPAS simulations. An isobaric lid with a vanishing vertical pressure velocity $\omega_p$ at the upper boundary leads to numerical Lamb waves with a model top height-dependent speed, as verified in the SE and FV3 simulations. The speed reduction with an isobaric lid is significant for low model tops: with an isothermal temperature of $T_0=275$ K, a model top height of 10 km leads to a numerical Lamb wave that is 25\% slower than the classical Lamb wave. The upper boundary condition also impacts the vertical structure of the numerical Lamb wave. With a rigid lid, the $\exp(2z/(7H))$ vertical profile is obtained for $u$ throughout the domain, whilst it is obeyed with an isobaric lid for all but the uppermost levels, where the vertical gradient of the zonal wind reduces to satisfy $du/dz=0$ at the upper boundary. Hence, the rigid lid provides a more faithful representation of the classical Lamb wave that exists in an isothermal atmosphere than an isobaric lid. \par
We also examined the impact of model parameters on the numerical Lamb wave. An increased horizontal resolution led to stronger Lamb wave amplitudes in all dynamical cores. In FV3, there is sensitivity to the dynamics time step, with a slower and less diffused wave for a smaller dynamics time step due to increased subcycling. We also showed that MPAS' time-adjusted acoustic filter is effective in damping the Lamb wave in this test. SE had the strongest numerical Lamb waves, and the exact reasons for this are left for future research. We are also interested in applying our Lamb wave test case to other dynamical cores, such as those with a Charney-Phillips vertical staggering or a different vertical coordinate. \par

\section*{Code Availability}
Simulations were performed using clones of the CAM GitHub repository, \url{https://github.com/ESCOMP/CAM}. The tag numbers used were \texttt{cam\_6\_4\_089} for SE, \texttt{cam\_6\_4\_095} for FV3, and \texttt{cam\_6\_4\_161} for MPAS. The CAM tag for MPAS includes the MPAS-standalone version 8.2. Tests with a later CAM tag \texttt{cam\_6\_4\_196} with the newer MPAS-standalone version 8.4 were also conducted which led to identical MPAS results. Plotting scripts, the hybrid coefficients for the hybrid pressure-based vertical coordinates, and a record of xmlchanges and namelist options for the CAM $1^\circ$ simulations can be viewed in the GitHub repository \url{https://github.com/nick4cone/LambWavePaper}. The namelist parameters include the settings for the dissipation mechanisms and their coefficients for each dynamical core. 

\section*{Acknowledgements}
This work was supported by the National Oceanic and Atmospheric Administration (NOAA) grant NA22OAR4320150 and the National Science Foundation grants 2332468 and 2608195. We would like to acknowledge high-performance computing support from the Derecho (doi:10.5065/qx9a-pg09) and Casper (https://ncar.pub/casper) systems provided by NSF NCAR, sponsored by the National Science Foundation.

\section*{Declaration of Generative AI and AI-assisted technologies in the writing process}
This manuscript was entirely written by the authors. AI tools of Claude Opus and UMGPT (from the University of Michigan) were used for feedback on the manuscript drafts, and the authors take full responsibility for the content of the publication.

\appendix
\section{Numerical Lamb wave speeds with an isobaric lid}
\label{append:cL_p_coord}
Models with an isobaric lid generate a Lamb wave with a speed that depends on the effective equivalent depth, $h^*$. Here, we discuss the computation of $h^*(z_{\text{T}})$ to obtain the numerical Lamb wave speed of $c^*=\sqrt{gh^*}$. This method follows the theory outlined in \textcite{de2020barotropic} for a bottom boundary condition of $w = 0$ and an upper boundary condition of $L_n(z) = 0$. For finite $z_{\text{T}}$, we have $c^*< c_{\text{L}}$, so the numerical Lamb wave in an isobaric lid model is slower than the classical Lamb wave. \par
Computing $h^*$ requires solving a semi-analytic problem that depends on the thickness of the layer, which in our case is the model top height of $z_{\text{T}}$. The semi-analytic problem falls into two regimes depending on $z_{\text{T}}$ relative to the scale height $H$: $z_{\text{T}} < 8H/3$ or $z_{\text{T}} > 8H/3$. For the critical case of $z_{\text{T}}=8H/3$, the equivalent depth has the analytical solution of $h^*=4 \kappa H$, leading to a numerical Lamb wave speed of $c^* = 2NH$, with $N$ the Brunt-V{\"a}is{\"a}l{\"a} frequency (\ref{eq:brunt_vais_freq}). \par
For $z_{\text{T}} \neq 8H/3$, we find the root of
\begin{equation}
    R(x) = 
    \begin{dcases}
       \tan(x) - \frac{4 \kappa H x}{z_{\text{T}} \bracked{1 - 2 \kappa + \bracfrac{2Hx}{z_{\text{T}}}^2}}, &~\text{if} ~z_{\text{T}} < \frac{8H}{3}, \\
       \tanh(x) - \frac{4 \kappa H x}{z_{\text{T}} \bracked{1 - 2 \kappa - \bracfrac{2Hx}{z_{\text{T}}}^2}}, &~\text{if} ~z_{\text{T}} > \frac{8H}{3}, 
    \end{dcases}
\end{equation}

\noindent with
\begin{equation}
    x = 
    \begin{dcases}
        \frac{z_{\text{T}}}{2 H} \sqrt{\frac{4 \kappa H}{h^*} - 1}, &~\text{if} ~z_{\text{T}} < \frac{8H}{3}, \\
        \frac{z_{\text{T}}}{2 H} \sqrt{1 - \frac{4 \kappa H}{h^*}}, &~\text{if} ~z_{\text{T}} > \frac{8H}{3},
    \end{dcases}
\end{equation}

\noindent which can be rearranged for effective equivalent heights,
\begin{equation}
    h^* = 
    \begin{dcases}
        \frac{4 \kappa H}{1 + \bracfrac{2Hx}{z_{\text{T}}}^2}, &~\text{if} ~z_{\text{T}} < \frac{8H}{3}, \\
        \frac{4 \kappa H}{1 - \bracfrac{2Hx}{z_{\text{T}}}^2}, &~\text{if} ~z_{\text{T}} > \frac{8H}{3}.
    \end{dcases}
\label{eq:h_star}
\end{equation}

In the $z_{\text{T}} < 8H/3$ case, we seek the smallest positive root (the roots at larger $x$ correspond to gravity waves, as discussed in Appendix \ref{append:gravity_wave_speed}), whilst there is only a single positive root when $z_{\text{T}} > 8H/3$. We use Brent's method to obtain the desired root of $R(x)$ within the intervals
\begin{equation}
    x \in
        \begin{dcases}
            \left[ \epsilon, \frac{\pi}{2} - \epsilon \right], &~\text{if} ~z_{\text{T}} < \frac{8H}{3}, \\
            \left[ \epsilon, \frac{z_{\text{T}} \sqrt{1 - 2 \kappa}}{2H} - \epsilon \right], &~\text{if} ~z_{\text{T}} > \frac{8H}{3}.
        \end{dcases}
    \label{eq:lamb_root_brackets}
\end{equation}

\noindent The introduction of an offset $\epsilon = 10^{-6}$ avoids obtaining the trivial solution at $x=0$ or hitting singularities in $R(x)$. Once the root $x$ is identified, the effective equivalent depth $h^*(x)$ (\ref{eq:h_star}) and isobaric lid Lamb wave speed of $c^* = \sqrt{gh^*}$ are computed.

\section{Numerical gravity wave speeds}
\label{append:gravity_wave_speed}
This appendix derives the phase speeds for the internal gravity waves. We then quote the speeds of the modes supported by a numerical model with $w=0$ boundary condition at the bottom surface, and either $w=0$ (rigid lid) or $\omega_p=0$ (isobaric lid) at the model top. \par
We begin with the general dispersion relation for acoustic-gravity waves in the presence of stratification, which requires solving the quartic equation
\begin{equation}
    \omega^4 - c_{\text{s}}^2 \omega^2 \bracked{k_{\text{h}}^2 + m^2 + \frac{1}{4H^2}} + c_{\text{s}}^2N^2 k_{\text{h}}^2 = 0,
\end{equation}

\noindent where the sound speed $c_{\text{s}}=\sqrt{\gamma R_{\text{d}} T_0}$ is equal to the classical Lamb wave speed (\ref{eq:lamb_speed}). Applying the quadratic formula to $\omega^2$ and taking the negative root gives the gravity wave dispersion relation of
\begin{equation}
    \omega_{\text{grav}}^2 = \frac{A c_{\text{s}}^2}{2}  (1 - \sqrt{1-\delta}),
\end{equation}

\noindent where we have introduced the quantities $A = k_{\text{h}}^2 + m^2 + 1/(4H^2)$ and $\delta = 4N^2k_{\text{h}}^2/(c_{\text{s}}^2A^2)$. The gravity wave dispersion relation can be rearranged as
\begin{equation}
    \omega_{\text{grav}}^2 =  \frac{Ac_{\text{s}}^2}{2} \frac{\delta}{1 + \sqrt{1 - \delta}} = \frac{2N^2k_{\text{h}}^2}{A(1 + \sqrt{1 - \delta})},
\end{equation}

\noindent to identify a horizontal phase speed of
\begin{equation}
    c_n(k_{\text{h}}) = \frac{\omega_{\text{grav}}}{k_{\text{h}}} =  N\sqrt{\frac{2}{A(1 + \sqrt{1-\delta})}}.
\end{equation}

Consider the long-wave limit of $k_{\text{h}} \rightarrow 0$, which means that $A \rightarrow m^2 + 1/(4H^2)$ and $\sqrt{1-\delta} \rightarrow 1$. Hence,
\begin{align}
    &~c_{n,\text{lw}} = \lim_{k_{\text{h}} \rightarrow 0} c_n (k_{\text{h}}) = \frac{N}{\sqrt{m^2 + 1/(4H^2)}}, \\
    \rightarrow &~c_{n,\text{lw}} = \frac{2NH}{\sqrt{1 + (2mH)^2}},
\end{align}

\noindent which is the expression given in (\ref{eq:grav_speeds}). \par

With a rigid lid, we have vertical wavenumbers of
\begin{equation}
    m_n = \frac{n\pi}{z_{\text{T}}}, 
\end{equation}

\noindent for integers $n \geq 1$. With an isobaric lid, there is no analytic formula directly for the effective wavenumbers $m_n^*$. Instead, following \textcite{de2020barotropic} we solve
\begin{equation}
    \tan(x_n) - \frac{4 \kappa H x_n}{z_{\text{T}} \bracked{1 - 2 \kappa + \bracfrac{2Hx_n}{z_{\text{T}}}^2}}=0
\label{eq:grav_speed_p_coord_Rx}
\end{equation}

\noindent for $x_n=m_n^* z_{\text{T}}$. Note that (\ref{eq:grav_speed_p_coord_Rx}) is the same problem as for the Lamb wave in the $z_{\text{T}} < 8H/3$ regime, as discussed in Appendix \ref{append:cL_p_coord}. Whereas for the Lamb wave we sought a root in the interval $(0, \pi/2)$ (\ref{eq:lamb_root_brackets}), the gravity wave mode $n$ is obtained from the root in the interval $x_n \in (n\pi, n\pi+\pi/2)$. As $x_n > n\pi$, $m_n^* > n\pi/z_{\text{T}} = m_n$, and so $c_{n,\text{lw}}^* < c_{n,\text{lw}}$ for all vertical modes $n$. This means that each gravity wave mode is slower with an isobaric lid compared to its rigid lid counterpart (e.g. Figure \ref{fig:analyt_lamb_grav_speed_with_T0}).\par

\section{Relationship between $L_n(z)$ and horizontal velocity}
\label{append:Ln_to_dUdz}
This appendix shows that the vertical gradient of the zonal wind is proportional to the vertical structure function $L_n(z)$; this explains that an isobaric lid causes $du/dz \rightarrow 0$ at the upper boundary as seen in Figure \ref{fig:vert_profile_p_pert}. A similar argument can be made for the meridional velocity, $v$. \par
We start with the phase relation for the vertical structure of the zonal winds $u_n(z)$, given in Eq.~(43) of \textcite{lindzen1969atmospheric}. Here $u_n(z)$, similar to $L_n(z)$, represents the functional dependency of $u$ on height when using separation of variables. The phase relation is expressed in terms of $y_n$, a change of variable for $L_n$ given by $y_n=L_n\exp(-\int{dz}/H)$. Applying this relationship, we obtain the phase relation in terms of $L_n$, 
\begin{equation}
    u_n = \frac{\gamma g h_n}{4a\Omega^2}\bigg(H\frac{dL_n}{dz}-L_n\bigg), \label{eq:u_phase}
\end{equation}

\noindent where $a$ is the radius of the Earth and $\Omega$ the rotation rate of the Earth. While the division by the rotation rate appears to cause a singularity when $\Omega\to 0$, it is simply the result of how \textcite{lindzen1969atmospheric} define constants in relation to the shallow water system. The same proportionality to the quantities within the parentheses holds for a nonrotating system as well. \par
By taking a derivative of (\ref{eq:u_phase}) with respect to height we find 
\begin{equation}
    \frac{\partial u_n}{\partial z}=\frac{\gamma g h_n}{4a\Omega^2}\bigg(H\frac{d^2L_n}{dz^2}+\bigg(\frac{dH}{dz}-1\bigg)\frac{dL_n}{dz}\bigg).
\end{equation}
The terms in parentheses can be rewritten using the vertical structure equation (Eq. 27 in \textcite{lindzen1969atmospheric}),
\begin{equation}
H\frac{d^2L_n}{dz^2}+\bigg(\frac{dH}{dz}-1\bigg)\frac{dL_n}{dz}+\frac{1}{h_n}\bigg(\frac{dH}{dz}+\kappa\bigg)L_n=0,
    \label{eq:struc_eqn}
\end{equation}
where the forcing terms have been set to zero to obtain a direct relationship to $L_n$,
\begin{equation}
    \frac{\partial u_n}{\partial z}=-\frac{\gamma g}{4a\Omega^2}\bigg(\frac{dH}{dz}+\kappa\bigg)L_n.
    \label{eq:dUdz_general}
\end{equation}
For an atmosphere with an isothermal base state, $H$ is a constant and so $\partial u_n/\partial z$ is directly proportional to $L_n$. Hence, in the case where $L_n=0$ at the model top, the vertical derivative of the horizontal winds approaches $0$ as well. This is shown in Figure \ref{fig:vert_profile_p_pert} with the flattening of the vertical gradient of $\ln(u)$ near the model top for the isobaric lid models of SE and FV3.

\printbibliography

\end{document}